\pdfoutput=1
\documentclass[12pt,twoside,journal,draftclsnofoot,onecolumn]{IEEEtran}

\usepackage{amsmath,amsfonts,bm,amssymb,psfrag,ifthen,color,subfigure}
\usepackage{mathrsfs}
\usepackage[justification=centering,font=footnotesize]{caption}
\usepackage{todonotes}

\allowdisplaybreaks[4]
\usepackage{gensymb}

\usepackage[dvips]{epsfig}
\usepackage[dvips]{graphicx}
\usepackage{amsmath,amsfonts,bm,amssymb,psfrag,ifthen,color,subfigure}
\usepackage{algpseudocode}
\usepackage{algorithm}
\usepackage{algorithmicx}
\usepackage{ifthen}
\usepackage{setspace}
\usepackage{cite}
\usepackage{url}
\usepackage{longtable}
\usepackage{stfloats}  
\usepackage{graphics,booktabs,color,subfigure}
\usepackage{float}
\usepackage{diagbox}

\usepackage[dvips]{epsfig}
\usepackage[dvips]{graphicx}
\usepackage{amsmath,amsfonts,bm,amssymb,psfrag,ifthen,color,subfigure,amsthm}
\usepackage{algpseudocode}
\usepackage{algorithm}
\usepackage{algorithmicx}
\usepackage{setspace}
\usepackage{cite}
\usepackage{url}
\usepackage{longtable}
\usepackage{stfloats}  
\usepackage{graphics,booktabs,color,subfigure}
\usepackage{float}
\usepackage{diagbox}
\usepackage{multirow}

\allowdisplaybreaks[4]

\newcommand{\tabincell}[2]{\begin{tabular}{@{}#1@{}}#2\end{tabular}}

\usepackage{mathrsfs}

\usepackage{graphicx}
\usepackage{makecell}
\usepackage{epstopdf}

\usepackage{graphicx}
\usepackage{todonotes} 

\begin{document}
	\title{ Semantic Feature Division Multiple Access for Multi-user Digital Interference Networks}

		\author{Shuai Ma,  Chuanhui Zhang, Bin Shen,  Youlong Wu,   Hang Li, Shiyin Li, Guangming Shi, and Naofal Al-Dhahir

 	\thanks{Shuai Ma is with Peng Cheng Laboratory, Shenzhen 518066, China (e-mail:  mash01@pcl.ac.cn).}

}
	\maketitle
	\begin{abstract}
		
	With the ever-increasing   user density and  quality of service (QoS)  demand, 5G networks with limited spectrum   resources are facing  massive access  challenges.
To address these  challenges,  in this paper, we
propose  a novel discrete semantic feature division multiple access (SFDMA) paradigm for   multi-user digital interference networks. Specifically, by   utilizing deep learning technology, SFDMA extracts   multi-user semantic information    into   discrete representations  in distinguishable semantic subspaces,  which enables multiple users to transmit simultaneously over the same time-frequency resources. Furthermore, based on a robust information bottleneck, we  design a   SFDMA based  multi-user digital semantic
interference network   for  inference tasks,   which can achieve    approximate orthogonal transmission.
Moreover, we propose a SFDMA based multi-user digital semantic
interference network   for image     reconstruction tasks, where  the discrete outputs of the  semantic encoders of the users are approximately orthogonal, which  significantly reduces multi-user interference.
Furthermore, we propose an   Alpha-Beta-Gamma (ABG)  formula  for  semantic communications, which is the  first    theoretical relationship between inference   accuracy  and transmission power.  Then,  we derive adaptive power control methods with closed-form expressions for  inference tasks. Extensive simulations verify the effectiveness and superiority of the proposed SFDMA.

	\end{abstract}
	
	\IEEEpeerreviewmaketitle

	\section{Introduction}
	 The explosive growth of mobile devices and  the emergence of numerous  intelligent applications, such as virtual reality (VR)\cite{Lai_2023_VR}, digital twin, multi-sense experience, and Metaverse\cite{Yu_2023_Metaverse}, present a new challenge to  fifth-generation (5G)    wireless networks:  massive access requirement with limited spectrum   resources. Due to the broadcast nature of wireless communications, the
	interference between multiple concurrent information flows is a major obstacle limiting multi-user network capacity,
	and it is even more severe in massive Internet of Things (IoT) networks\cite{Luo_2022_SOO}.
	Therefore, finding an efficient  multiple access (MA) scheme to manage multi-user interference
	is critical for  fulfilling  the ultimate goal of  ultra-dense  access.\par
	
	The crux of the MA problem of  wireless networks is to
	allocate  radio resources to   multiple users\cite{Saito_2013}. The existing   various MA schemes
	  can be categorized into orthogonal multiple access (OMA) schemes and non-orthogonal multiple access (NOMA)\cite{Mao_RSMA_2018} schemes.
	Specifically, for OMA schemes,  the allocated resources are  orthogonal in the frequency, time, coding  or spatial domains, which restricts the number of users accessing the network due to the limited resources.  To further enhance   networks capacity, MA has progressed towards
	NOMA.
	By employing   superposition coding  at the transmitter   and successive interference cancellation (SIC)\cite{Song_2017_resource} at the receivers,
	NOMA manages multi-user interference by forcing (at least) one user to successfully decode messages (and remove interference) of other users.
	 However, the complexity of SIC is relatively high, and the design of the receiver is challenging.\par
	
	Thus, it is imperative to  develop novel effective multi-user interference management paradigms to boost the area spectral efficiency
	and fulfill the vision of massive intelligent connectivity.
	Recently, semantic communication  has drawn great attention due to its capability of significantly reducing the amount of  the transmitted data\cite{Dai_CBTB_2023}.
	There has been an increasing volume of works in semantic communication. The existing works can be categorized either by the applications or by the information sources.
		From the perspective of applications,  the existing
	research works can be categorized  into two classes:  task-oriented communication and  data reconstruction.
		\begin{itemize}
		\item
		For task-oriented communications, based on a   retrieval-oriented deep image compression scheme,the authors in \cite{Jankowski_2021} proposed      both digital and analog JSCC schemes for wireless image retrieval at the wireless edge.
		Inspired by  information bottleneck (IB) theory, the authors in  \cite{Shao_2022} investigated the rate-distortion tradeoff between
		the encoded feature's informativeness and the inference performance  for task-oriented semantic communications.
		Furthermore,  by  exploiting a robust information bottleneck (RIB) framework, the authors in \cite{Xie_2023}
		studied the tradeoff between the informativeness of the encoded representation and the robustness against channel variation in task-oriented communications.

		\item 	
	For data reconstruction tasks,  the authors of \cite{Bourtsoulatze_2019} put forward a deep joint source-channel coding (JSCC)  architecture, where the encoder and   decoder  are parameterized by CNNs  to minimize the average mean squared error (MSE) of the reconstructed image.
	By combining  MSE and structural similarity index matrix (SSIM) as the loss function, an autoencoder-based JSCC scheme was proposed in \cite{Yan_DLA_2021}  to
	explore both pixel-wise and structural features of the images. By  leveraging nonlinear transform coding, a  nonlinear transform source-channel coding
	(NTSCC) architecture was devised in  \cite{Dai_2022} to minimize the end-to-end images transmission rate-distortion performance.
	In \cite{Huang_TSC_2023}, a RL-based adaptive  semantic information coding scheme was designed  for  rate-semantic-aware   image transmission.
	
	\end{itemize}

	
	From the perspective of information sources, the existing works can also be divided into: text, speech, image and video.
    By integrating hybrid automatic repeat request (HARQ) and Reed Solomon (RS) channel coding,  an end-to-end text semantic coding  architecture was developed in \cite{Jiang_2022} for sentence semantic transmission with varying lengths. 	
	To reflect the semantic fidelity of the model, a deep learning-based joint source-channel coding semantic communication system architecture using sentence-level semantic information is proposed in \cite{Tang_Text_2023}, and a new semantic similarity measure method is proposed to evaluate the semantic fidelity.
	By utilizing an attention mechanism with a squeeze-and-excitation (SE) network, a DL-enabled semantic communication system was proposed \cite{Weng_2021} to  extract the semantic information  of   speech signals and transmit over various channel conditions.
	Using an attention-based soft alignment module and a redundancy removal module, a highly semantically focused communication system is proposed in \cite{Han_SPC_2023}, which only extracts text-related semantic features and removes semantically irrelevant features for speech-to-text transmission and speech-to-speech transmission
	By introducing  the concept of semantic slice-models (SeSM), a layer-based semantic coding communication system was designed  in  \cite{Dong_SCS_2023} for   transmission and recovery of image semantic information.
	By using only key point delivery to represent facial expression motion, a semantic transmission framework for video conferencing is proposed in \cite{Jiang_WS_2023}, which introduces  an incremental redundancy hybrid automatic repeat-request framework and combines a new semantic error detector to deal with different channel environments in video conferencing systems.
	
	
Note that, most of the aforementioned works focus on   single-user semantic communication scenarios (i.e., the one-to-one
	or point-to-point communication) without multi-user interference.
However, multi-user semantic communication networks,  as the most common communications scenario,
 are still     in their infancy.
Different from the  single-user semantic communication  scenarios, multi-user interference is a  critical bottleneck for improving     capacity  of multi-user communication networks.
	Based on deep neural network (DNN), the authors of \cite{Hu_OTM_2022}  designed a semantic communication system for the broadcast scenario, where two receivers with a semantic recognizer distinguish	 positive  and negative sentences.
	For visual question answering (VQA) tasks, a multi-user task-oriented communication  system  was developed in \cite{Xie_VQA_2022}  by using multiple antennas linear minimum mean-squared error (L-MMSE) detector and joint source channel decoder to mitigate the effects of channel distortion and inter-user interference. Moreover, in \cite{Zhang_MUS_2022}, a multi-user semantic communication system is studied to execute object-identification tasks, where correlated source data among different users is transmitted via a shared channel.
	In \cite{Luo_MM_2022},  the authors proposed a  multi-modal information fusion scheme for multi-user semantic communications,  where the wireless channel acts as a medium to fuse multi-modal data where a receiver   retrieves semantic information without the need to perform multiuser signal detection.
By using   attention and residual structure modules,
 the authors in \cite{Zhang_TCCN_2023}  developed  a DL-based multiple access
method for continuous
semantic symbols transmission in  image reconstruction
tasks.
However, in the  existing multi-user semantic communication works\cite{Hu_OTM_2022,Xie_VQA_2022,Zhang_MUS_2022,Luo_MM_2022,Zhang_TCCN_2023}, JSCC  directly maps
the source data into continuous channel input symbols, which is incompatible
  with   current digital communication systems.
More specifically,  the direct transmission of continuous
feature representations requires analog
modulation or a full-resolution constellation, which brings
huge burdens for resource-constrained radio frequency
systems.

Moreover, the performance of semantic communication depends on the   transmit power. However,  the existing semantic communication performance measurements  are  end-to-end, such as  classification accuracy for inference tasks, which have not yet established a relationship with transmit power. The main reason is that  DL based semantic encoders  are  generally highly complex nonlinear functions, and it is hard to derive analytical relationships between end-to-end performance measures and transmit power. Therefore,   there is no theoretical basis for  adaptive power control for  semantic communications, which leads to performance degradation   in random fading channels.


To address the above challenges,  we explore a new resource domain: semantic feature domain by utilizing DL technology, and propose a semantic feature division multiple access (SFDMA) scheme, which is different from the existing multiple access schemes based on the time, frequency, code, spatial,  or power domains.
Specifically, the DL based semantic encoders extract
  semantic information into discrete semantic feature representations
 in distinguishable semantic subspaces, and the discrete semantic    feature vectors of the multiple users are approximately   orthogonal to each other.
 Furthermore, based on the SDFMA scheme,  we  design multi-user digital semantic interference networks     for inference tasks and image reconstruction tasks. Then,  we  establish the relationship between  inference   accuracy   and transmission power.
 The main contributions of this paper can be summarized as follows:
\begin{itemize}
	\item 	
	By exploring the feature domain,   an  SFDMA  scheme  is proposed  for multi-user digital semantic communication networks, where the discrete encoded features of multiple users  are   in the distinguishable feature domain.
	Specifically, by utilizing   computation capability,  the transmitter  encodes   the user information into discrete semantic feature representations  in the feature domain, where the discrete semantic features   of   different users are   approximately orthogonal. Then,  the receiver extracts and decodes the intended semantic information. Since the semantic features of multiple users are   in the distinguishable feature domain,  multi-user  interference is significantly reduced.


	\item Based on the proposed SFDMA framework, we design a digital multi-user semantic interference network  with inference tasks.
	To achieve  the informativeness-robustness-multiuser interference tradeoff,   the proposed RIB based SFDMA scheme formulates the informativeness-robustness-multiuser interference tradeoff in the encoded representation and aims at maximizing the coded redundancy to improve  robustness, while restricting the interference for other users and  retaining sufficient information for the   inference tasks. Due to the computational intractability of mutual information, we derive the tractable variational upper bound of the RIB objective by utilizing the variational approximation technique.  The proposed RIB based SFDMA scheme can  realize nearly orthogonal and high-level transmission of user semantic features, while protecting user semantic information from being decoded by other users.
	
	\item Based on the   Swin Transformer,  we develop
	both centralized and distributed coordinated SFDMA for   image  reconstruction tasks in multi-user interference
	networks. The proposed   semantic encoders encode the semantic information of  each user into  distinguishable semantic subspaces, and the extracted semantic features are  approximately orthogonal, which  significantly reduces multi-user interference.
	More importantly, SFDMA can protect privacy of users' semantic information, in which
	the semantic information can only be decoded by the corresponding semantic receiver and  cannot be decoded  by other receivers.

\item  Furthermore, we  reveal  the relationship between end-to-end performance measurements: classification accuracy  and transmission power, which can be approximately fitted to an Alpha-Beta-Gamma (ABG) functions. To the best of our knowledge, this is the first theoretical   expression between   inference   accuracy  and transmission power for  semantic communications.
    Based on the ABG function,  we propose  adaptive power control methods with closed-form expression for  inference tasks. The adaptive power control method   can   effectively guarantee the quality of   service (QoS)  semantic communication in random fading channels.

\end{itemize}

The rest of this paper is organized as follows. Section II introduces an SFDMA scheme for multi-user semantic interference networks. The  multiuser semantic  interference network with inference task is described  in Section III,  presents. Section IV presents the multi-user semantic interference network for image reconstruction.  In Section V, we propose  the adaptive power control scheme for random fading channels. In Section VI, the experimental results and analysis are presented. Finally, Section VII concludes the paper.

For our notation, we denote random variables by capital letters (e.g. $X$) and their realizations by  boldfaced lowercase letters (e.g. $\mathbf{u}_i$). The notations $(.)^H$, $(.)^{-1}$
and $\mathbb{E}[.]$ represent the transpose, inverse and expectation of a matrix, respectively. 
The notations used in this paper are summarized in  TABLE \ref{table1}.
{\small
\begin{table}[htbp]
	\caption{Summary of Key Notations}
	\label{table1}
	\centering
	\begin{tabular}{|m{1.5cm}<{\centering}|m{7cm}|}
		\hline
		\rule{0pt}{8pt}Notations  &    Meanings \\ \hline
		
		\rule{0pt}{7.5pt}$\mathbf{s}_i$ &  \tabincell{c}{Input data of TX $i$} \\ \hline
		
		\rule{0pt}{7.5pt}$\mathbf{u}_i $ &  Semantic information of TX $i$  \\ \hline
		
		\rule{0pt}{7.5pt}$\mathbf{x}_i$ &  Encoded semantic feature of TX $i$ \\ \hline
		
		\rule{0pt}{7.5pt}${{\cal T}_i}$ &   Semantic feature  subspace of TX $i$\\ \hline	
		
		\rule{0pt}{7.5pt}${\mathbf{g}_{i,j}}$ &  \tabincell{c}{ The channel gain from TX $j$ to RX
			$i$}\\ \hline
		\rule{0pt}{7.5pt}  ${\psi_i}$  &The parameters set of   semantic encoder of TX $i$  \\ \hline

 		\rule{0pt}{7.5pt}  $f_{\psi_i}\left(  \cdot  \right)$  & Semantic encoder  of TX $i$   \\ \hline
		\rule{0pt}{7.5pt}  ${\theta_i}$  &  The parameters set of   semantic decoder  of RX $i$  \\ \hline

		\rule{0pt}{7.5pt}  $f_{\theta_i}\left(  \cdot  \right)$  & Semantic decoder of RX $i$   \\ \hline
		
		\rule{0pt}{7.5pt} $ Q(.)$  &  \tabincell{c}{Quantizer}  \\ \hline
		\rule{0pt}{7.5pt} $\Theta_{m}(.)$   &  \tabincell{c}{Modulator}  \\ \hline
  		\rule{0pt}{7.5pt} $ d $  &  \tabincell{c}{Number of quantization bits}  \\ \hline
	\end{tabular}
\end{table}
}

	
	\section{   The SFDMA for  Multi-user   Semantic Interference Networks}

		\begin{figure}[!t]
		\centering
		\includegraphics[width=14cm]{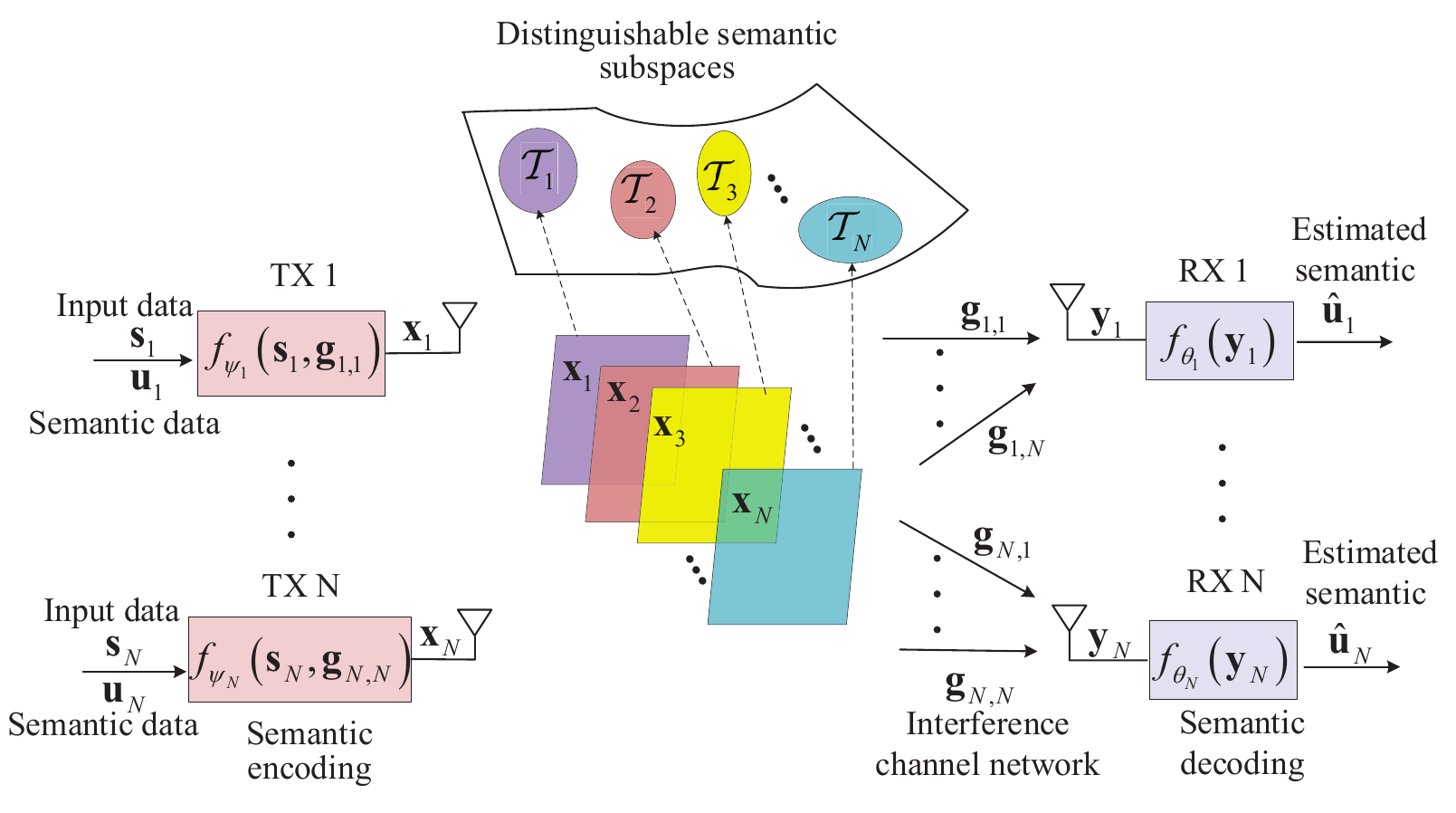}
		\caption{ SFDMA based multi-user semantic interference networks.}
		\label{fig1}
	\end{figure}
	Consider a typical multi-user  interference channel  with $N$ transmission
	pairs, as illustrated in Fig. \ref{fig1}, where each transmission
	pair includes one semantic transmitter (TX) and one semantic receiver (RX).
	Let ${{{\bf{s}}_i}}$ denote the  data with   implicit
	semantic information ${{{\bf{u}}_i}}$  of the $i$th transmission pair.
In the semantic interference network, the  $N$ TXs adopt the SFDMA scheme to
	   simultaneously transmit information in the same time-frequency
	resources.

	Specifically, via the semantic encoder $f_{\psi_i}\left(  \cdot  \right)$, the   input data ${{{\bf{s}}_i}}$ is encoded  to the semantic feature ${{{\bf{x}}_i}}$ in   the semantic features subspace ${{T_i}}$ where
	$\left\{ {{\cal T}_i} \right\}_{i = 1}^N $ as follows
	\begin{align}
	{{\bf{x}}_i} = {f_{{\psi _i}}}\left( {{{\bf{s}}_i},{\mathbf{g}_{i,i}}} \right), {i \in \left\{ {1,...,N} \right\}},
	\end{align}
	where ${\bf{g}}_{i,i} $ denotes the channel gain from TX $i$ to RX  $i$.
	Moreover, to avoid multi-user interference, the semantic features subspaces $\left\{ {{\cal T}_i} \right\}_{i = 1}^N $ satisfy the following conditions
	\begin{align}
	\left\{ \begin{array}{l}{{\bf{x}}_i} \in {{\cal T}_i},{{\bf{x}}_i} \notin {{\cal T}_j},\forall i \ne j,\\
	{{\cal T}_i} \cap {{\cal T}_j}{\rm{ = }}\emptyset, \forall i \ne j.
	\end{array} \right.
	\end{align}
	where
	$i, j \in \left\{ {1,...,N} \right\}$.	
	In other words,  the semantic encoder  separates the semantic feature space $T$ into multiple distinct semantic feature subspaces $\left\{ {{{\cal T}_i}} \right\}_{i = 1}^N $, where  the  multiple user signals  are encoded and  transmitted  in the separated semantic feature subspaces ${{\bf{x}}_i} \in {{\cal T}_i}$, ${{\bf{x}}_j} \notin {{\cal T}_i},\forall j \ne i$.

For practical applications, it is hard to achieve perfect separation of feature subspaces. Therefore, the semantic feature vectors   are approximately orthogonal, and the inner product tends to 0, i.e.,
	\begin{align}{\bf{x}}_i^H{{\bf{x}}_j} \to 0,\forall i \ne j.
\end{align}

	Then,  the received signal of      RX $i$ is given as
	\begin{align}\label{777}
	{{\bf{y}}_i} = \underbrace {{{\bf{g}}_{i,i}} \odot {{\bf{x}}_i}}_{{\rm{desired\ singnal}}} + \underbrace { \sum\limits_{j = 1,j \ne i}^N{{\bf{g}}_{i,j}} \odot  {{{\bf{x}}_j}} }_{{\rm{interference}}} + \underbrace {{{\bf{n}}_i}}_{{\rm{noise}}},\end{align}
	where $ \odot $ is the Hadamard product\cite{Horn_1990_hadamard}, which denotes element-wise multiplication. The term  ${{{\bf{g}}_{i,i}} \odot  {{\bf{x}}_i}}$ is the desired signal of the  $i$th RX, $ \sum\limits_{j = 1,j \ne i}^N{{{\bf{g}}_{i,j}} \odot{{{\bf{x}}_j}} }$ is the multi-user semantic interference, and ${{\bf{n}}_i}\sim\mathcal{CN}\left(0,{\sigma_i}^2{\bf{I}}\right)$ is received noise of    RX $i$.
		Furthermore, by applying semantic decoder $f_{\theta_i}\left(  \cdot  \right)$,   RX  $i$ decodes the received signal ${{\bf{y}}_i}$, and  obtains the estimated semantic information ${\widehat {\bf{u}}_i}$ as follows
	\begin{align}\label{999}
	{\widehat {\bf{u}}_i} = {f_{{\theta _i}}}\left( {{{{\bf{y}}_i}}} \right),i \in \left\{ {1,...,N} \right\},
	\end{align}
	where the multi-user interference can be eliminated by the semantic decoder of RX  $i$, i.e., ${f_{{\theta _i}}}\left( \sum\limits_{j = 1,j \ne i}^N  {{{\bf{g}}_{i,j}} \odot {{{\bf{x}}_j}} } \right) = 0$.
	
	By the   intended signal ${{\bf{x}}_i}$  and  the multi-user interference are encoding into separated  feature subspaces, i.e., $\sum\limits_{j = 1,j \ne i}^N {{{\bf{x}}_j}}  \notin {{\cal T}_i}$, the  decoder  of each RX can effectively eliminates multi-user interference and decodes  the intended semantic information.
	
In the following, we will investigate the SFDMA based multiuser semantic interference
networks design for inference task  and  image reconstruction, respectively, and develop an adaptive power control scheme for semantic interference
networks.

	
	\section{Multi-user Semantic Interference  Network with Inference Tasks}

\begin{figure}[!t]
		\centering
		\includegraphics[width=14cm]{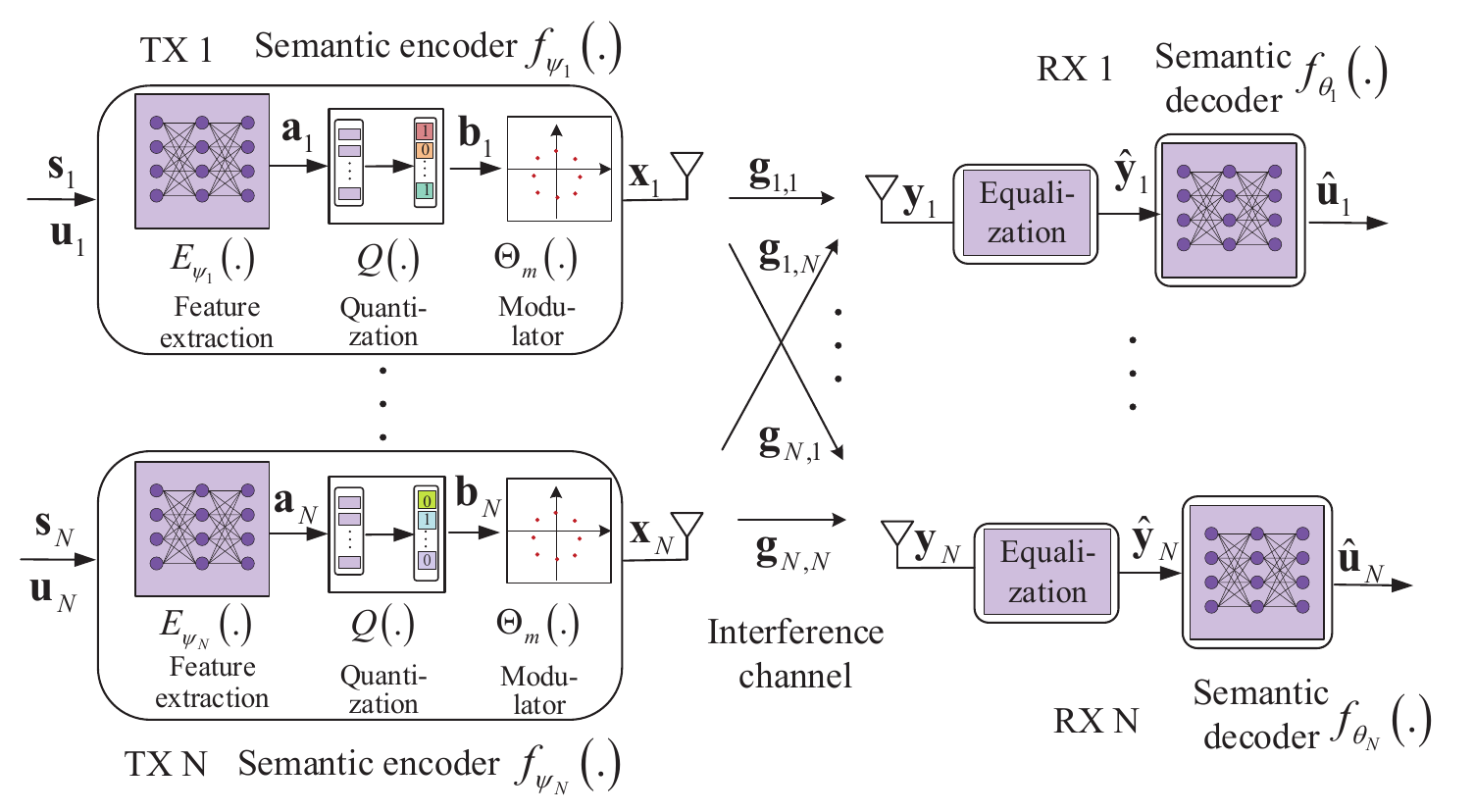}
		\caption{SFDMA based   semantic interference  network with inference tasks.}
		\label{fig2}
	\end{figure}

	In this section, we investigate a SFDMA based multi-user semantic interference  network with inference tasks.
	As shown in Fig. \ref{fig2},  by using a feature extraction network  ${{E_{{\psi _i}}}\left( . \right)}$, TX $i$  extracts semantic feature  $\mathbf{a}_i$ from the source data $\mathbf{s}_i$  as follows
	\begin{align}
	\mathbf{a}_{i}=E_{\psi_{i}}\left(\mathbf{s}_{i}\right),i\in\left\{1,...,N\right\}.
	\end{align}
	
	Note that, the  feature representations $\mathbf{a}_i$  is continuous, and
	the direct transmission of
	continuous feature representation  needs to be modulated with
	analog modulation or a full-resolution constellation, which
	brings huge burdens for a resource-constrained transmitter and
	poses implementation challenges on the current radio frequency
	(RF) systems. Moreover, modern mobile systems are based on digital
	modulation. Thus, in order to be compatible with current digital
	communication systems,  we employ a    digital modulation method for multi-user semantic interference communication networks.
    Specifically,     by applying a linear layer  ${\rm{Linear}(.)}$ and sign function ${\rm{Sign}(.)}$,  the feature vector $\mathbf{a}_{i}$ is  quantized into $d$ bits, i.e.,
	\begin{align}
	{{\bf{b}}_i} = {\rm{Sign}}\left( {{\rm{Linear}}({{\bf{a}}_i})} \right). \end{align}
	
	Note that, the quantization process  of  direct quantization can be formulated  as
	\begin{align}
	\mathbf{b}_i=Q\left(\mathbf{a}_i\right),~i\in\{1,..,N\}.
	\end{align}

	Furthermore, the quantized  feature representations $\mathbf{b}_i$ is  digitally modulated by  a modulator $\Theta_{m}$ and normalized to  $\mathbf{x}_{i}$ as follows
	\begin{align}
	\mathbf{x}_{i}={\rm{Norm}}\left(\Theta_{m}\left(\mathbf{b}_{i}\right)\right),i\in\{1,..,N\}.
	\end{align}
	
	In summary,
	the  semantic encoder  $f_{\psi_{i}}\left(  \cdot  \right)$ includes semantic feature extraction, quantization and modulation, i.e.,
	\begin{align}
	\mathbf{x}_i=f_{\psi_i}\left(\mathbf{s}_i\right),i\in\left\{1,...,N\right\},
	\end{align}
	where, $f_{\psi_i}\left(.\right)$ represents the semantic encoder of the TX  $i$,  and ${\psi_i}$ is the encoder parameter.

	Moreover,  the semantic encoder $f_{\psi_{i}}\left(.\right)$ can be modeled as a factorial categorical distribution with a probability mass function
	\begin{align}
	 p_{\psi_i}\left(\mathbf{x}_i\mid\mathbf{s}_i\right)=\prod_{r=1}^dp_{\psi_i}\left(x_{i,r}\mid\mathbf{s}_i\right).
	\end{align}

	Then, as shown in Fig. \ref{fig2},  the $N$ TXs simultaneously transmit    signals $\mathbf{x}_i$ to $N$ RXs,
	where  $\mathbf{g}_{i,i}$ is the channel gain from  TX $i$  to RX $i$,  $\mathbf{g}_{i,j}$ is the channel gain from  TX $j$    to  RX $i$, and  ${p_i}$ is the transmission power of the TX $i$.
	The received signal of   the  RX $i$  $\mathbf{y}_{i},i\in\left\{1,...,N\right\}$ is  given by
	\begin{align}\label{y11}
	{{\bf{y}}_i} = {{\bf{g}}_{i,i}} \odot {\sqrt{p_i}} {{\bf{x}}_i} +  \sum\limits_{j = 1,j \ne i}^N {{\bf{g}}_{i,j}} \odot{\sqrt{p_j}} {{{\bf{x}}_j}}  + {{\bf{n}}_i},
	\end{align}
	where   ${{\bf{n}}_i}$ denotes the received additive white Gaussian noise of the RX  $i$, i.e., ${{\bf{n}}_i}\sim\mathcal{CN}\left(0,{\sigma_i}^2{\bf{I}}\right)$.
	
    Furthermore, the channel equalization operation is carried out. We assume that the channel state information is perfectly known, and therefore, the received signal ${{{\bf{\hat y}}}_i}$ after equalization \cite{Xie_VQA_2022}can be transformed to
    \begin{subequations}\label{y1}
    	\begin{align}
     &{{{\bf{\hat y}}}_i} = {\left( {{\bf{g}}_{i,i}^H{{\bf{g}}_{i,i}}} \right)^{ -   1}}{\bf{g}}_{i,i}^H{{\bf{y}}_i}\\
    & = \sqrt {{p_i}} {{\bf{x}}_i} +  \sum\limits_{j = 1,j  \ne i}^N{\bf{g}}_{i,i}^{ - 1}{{\bf{g}}_{i,j}} \odot{\sqrt {{p_j}} } {{\bf{x}}_j} + {\bf{g}}_{i,i}^{ - 1}{{\bf{n}}_i}.
    	\end{align}
    \end{subequations}

	Finally,  the received signal ${{{\bf{\hat y}}}_i}$ is passed through semantic decoder $f_{\theta_{i}}\left(  \cdot  \right)$ to output the inference result ${\hat {\bf{u}}_i}$, i.e.,
	\begin{align}
	\mathbf{\hat{u}}_i=f_{\theta_i}\left(\mathbf{\hat y}_i\right),i\in\left\{1,...,N\right\},
	\end{align}
	where $f_{\theta_{i}}\left(  \cdot  \right)$ represents the semantic decoder of  RX  $i$,  and   ${\theta_i}$ denotes the  parameter set of the semantic  decoder network.

	\subsection{ Problem formulation}
	
	For the multi-user semantic interference  network, if the  TX  $i$ transmits more bits of
	information, it will improve the inference accuracy of
	the  RX  $i$ for decoding the semantic information $\mathbf{u}_i$, but
	it will also increase the transmission load of the  TX  $i$, and
	more importantly, it will increase interference for other receivers.
	The high interference will reduce the  decoding accuracy
	of other  receivers. Therefore, there is  a trade-off between the amount of transmitted information,
	interference between users, and inference accuracy for  multi-user semantic interference
	network.
	To achieve a compromise among the three, we propose
	an efficient semantic encoding and decoding scheme for the
	multi-user  semantic interference network based on the robust information bottleneck (RIB)\cite{Xie_2023}.
	
	Specifically, the semantic encoder of the  TX  $i$  $p_{\psi_i}\left(\mathbf{x}_i\mid\mathbf{s}_i\right)$     extracts semantic
	information about the target $\mathbf{u}_i$ while ignoring irrelevant information
	in the given transmit data $\mathbf{s}_i$. From a data compression
	perspective, the optimal discrete output variable $\mathbf{x}_i$ is the
	minimum sufficient statistic contained in the data $\mathbf{s}_i$  about
	the target $\mathbf{u}_i$. Another fundamental goal of communication
	system design is to maximize the communication transmission
	rate, which is achieved by maximizing the mutual information
	between the transmitted discrete output variable $\mathbf{x}_i$ and the
	received variable $\mathbf{y}_i$. Therefore, based on RIB principle, the robust information bottleneck for multi-user  semantic interference  network
	can be mathematically formulated as follows
	\begin{align}\label{IB}
	\mathop {\min }\limits_{\left\{ {{p_{{\psi _i}}}\left( {\mathbf{x}_i\mid \mathbf{s}_i} \right)} \right\}_{i = 1}^N} \sum\limits_{i = 1}^N  - I\left( {{U_i};{Y_i}} \right) - {\lambda _i}\Big[ {I\left( {{X_i};{Y_i}} \right) - I\left( {{S_i};{Y_i}} \right)} \Big],
	\end{align}
	where
	${I\left( {{U_i};{Y_i}} \right)}$ represents the correlation between the
	received signal ${{Y_i}}$ and the target ${{U_i}}$,    ${I\left( {{X_i};{Y_i}} \right)}$ denotes the correlation between the
	received signal ${{Y_i}}$ and the transmitted signal ${{X_i}}$, and ${I\left( {{S_i};{Y_i}} \right)}$
	denotes the correlation between the
	received signal ${{Y_i}}$ and the transmitted signal ${{S_i}}$. The parameter ${{\lambda _i}}$ controls the tradeoff
	between inference performance and model robustness.
	To develop robust semantic encoders ${\left\{ {{p_{{\psi _i}}}\left( {{x_i}\mid {u_i}} \right)} \right\}_{i = 1}^N}$, the
	optimization problem \eqref{IB} can be equivalently reformulated as
	follows	
	{\small
		\begin{subequations}\label{121}
		\begin{align}
		 &L_{\mathrm{RIB}}\left(\left\{\psi_{i}\right\}_{i=1}^{N}\right)=\sum_{i=1}^{N}-I\left(U_{i};Y_{i}\right)-\lambda_{i}\Big[I\left(X_{i};Y_{i}\right)-I\left(S_{i};Y_{i}\right)\Big] \\
		 &=\sum_{i=1}^N\mathbb{E}_{p(\mathbf{s}_i,\mathbf{u}_i)}\bigg\{\mathbb{E}_{p_{\psi_i}(\mathbf{y}_i|\mathbf{s}_i)}\Big[-\log p_{\psi_i}\left(\mathbf{u}_i\mid\mathbf{y}_i\right)\Big]\nonumber\\
		 &+\lambda_i\mathbb{E}_{P_{\psi_i}(\mathbf{x}_i|\mathbf{s}_i)}\Big[H\left(Y_i\mid\mathbf{x}_i\right)\Big]-\lambda_iH_{\psi_i}\big(Y_i\mid\mathbf{s}_i\big)\bigg\}-H\big(U_i\big) \\
		 &=\sum_{i=1}^N\mathbb{E}_{p(\mathbf{s}_i,\mathbf{u}_i)}\bigg\{\mathbb{E}_{p_{\psi_i}\left(\mathbf{y}_i|\mathbf{s}_i\right)}\Big[-\log p_{\psi_i}\left(\mathbf{u}_i\mid\mathbf{y}_i\right)\Big] \nonumber \\
		 &+\lambda_i\mathbb{E}_{P_{\psi_i(\mathbf{x}_i|\mathbf{s}_i)}}\Big[H\left(Y_i\mid\mathbf{x}_i\right)\Big]-\lambda_iH_{\psi_i}\left(Y_i\mid\mathbf{s}_i\right)\bigg\},\label{122}
		\end{align}
	\end{subequations}}where ${\lambda_{i}>0}$ is an adjustable weight factor, \eqref{122}  holds because the constant term ${H(U_{i})}$ is ignored in the  optimization. Specifically, $I\left(U_i;Y_i\right)$ denotes the information about the target $U_{i}$ that is held in $Y_{i}$. $I\left(S_i;Y_i\right)$ denotes the total amount of information encoded in $Y_{i}$ of $S_{i}$.
	
	Note that, the calculation of the posterior ${{p_{{\psi _i}}}\left( {{{\bf{u}}_i}\mid {{\bf{y}}_i}} \right)}$ in \eqref{122} involves high-dimensional integrals, i.e.,
	\begin{align}
	p_{\psi_i}\left(\mathbf{u}_i\mid\mathbf{y}_i\right)=\frac{\int p_{\psi_i}\left(\mathbf{u}_i,\mathbf{s}_i\right)p_{\psi_i}\left(\mathbf{y}_i\mid\mathbf{u}_i\right)d\mathbf{s}_i}{p_{\psi_i}\left(\mathbf{y}_i\right)},
	\end{align}
	which is intractable.

	To address this challenge, we exploit a variational  Bayesian\cite{fox_2012_tutorial} approach to approximate $p_{\psi_i}\left(\mathbf{u}_i\mid\mathbf{y}_i\right)$ by a variational distribution $p_{\psi_i}\left(\mathbf{u}_i\mid\mathbf{y}_i\right)$, i.e., ${q_{{\theta _i}}}\left( {{{\bf{u}}_i}\mid {{\bf{y}}_i}} \right) \approx {p_{{\psi _i}}}\left( {{{\bf{u}}_i}\mid {{\bf{y}}_i}} \right)$, where $\theta_{i}$ represents the learnable parameters of the neural network at the  RX  $i$.
	Specifically, the upper bound of the the first term of  \eqref{122} is given as
	\begin{subequations}
		\begin{align}
		 &\mathbb{E}_{p\left(\mathbf{s}_i,\mathbf{u}_i\right)}\bigg\{\mathbb{E}_{p_{\psi_i}(\mathbf{y}_i|\mathbf{s}_i)}\Big[-\log p_{\psi_i}\left(\mathbf{u}_i\mid\mathbf{y}_i\right)\Big]\bigg\} \\
		 &=\mathbb{E}_{p\left(\mathbf{s}_i,\mathbf{u}_i\right)}\bigg\{\mathbb{E}_{p_{\psi_i}(\mathbf{y}_i|\mathbf{s}_i)}\Big[-\log q_{\theta_i}\left(\mathbf{u}_i\mid\mathbf{y}_i\right)\Big]\bigg\}\nonumber \\
		 &-\underbrace{\mathbb{E}_{p_{\psi_i}(\mathbf{y}_i)}\left\{\mathbb{E}_{p_{\psi_i}(\mathbf{u}_i|\mathbf{y}_i)}\left[\log\frac{p_{\psi_i}\left(\mathbf{u}_i\mid\mathbf{y}_i\right)}{q_{\theta_i}\left(\mathbf{u}_i\mid\mathbf{y}_i\right)}\right]\right\}}_{D_{KL}\left(p_{\psi_i}\left(\mathbf{u}_i|\mathbf{y}_i\right)\|q_{\theta_i}\left(\mathbf{u}_i|\mathbf{y}_i\right)\right)\geq0} \\
		 &{\leq}\mathbb{E}_{p(\mathbf{s}_{i},\mathbf{u}_{i})}\bigg\{\mathbb{E}_{p_{\psi_{i}}(\mathbf{y}_{i}|\mathbf{s}_{i})}\Big[-\log q_{\theta_{i}}\left(\mathbf{u}_{i}\mid\mathbf{y}_{i}\right)\Big]\bigg\},\label{133}
		\end{align}
	\end{subequations}
	where inequality \eqref{133} holds since  $D_{KL}\left(p_{\psi_i}(\mathbf{u}_i|\mathbf{y}_i)\|q_{\theta_i}(\mathbf{u}_i|\mathbf{y}_i)\right)\geq 0$.
	Moreover, since $Y_{i}$ is the transmission representation corrupted by additional Gaussian noise and other interferences,  the entropy   $H_{\psi_{i}}\left(Y_{i}\mid\mathbf{s}_{i}\right)$ is   lower bounded by
	\begin{align}
	H_{\psi_{i}}\left(Y_{i}\mid\mathbf{s}_{i}\right)\geq H_{\psi_{i}}\left(X_{i}\mid\mathbf{s}_{i}\right).
	\end{align}
	Therefore,  the objective function in \eqref{121} can be reformulated as
    \begin{equation}
	\begin{aligned}
	&L_{\mathrm{RIB}}\left(\left\{\psi_{i}\right\}_{i=1}^{N}\right)\leq L_{\mathrm{VRIB}}\left(\left\{\psi_{i},\theta_{i}\right\}_{i=1}^{N}\right) \\
	 &\triangleq\sum_{i=1}^N\mathbb{E}_{p(\mathbf{s}_i,\mathbf{u}_i)}\bigg\{\mathbb{E}_{p_{\psi_i}(\mathbf{y}_i|\mathbf{s}_i)}\Big[-\log q_{\theta_i}\big(\mathbf{u}_i|\mathbf{y}_i\big)\Big] \\
	 &+\lambda_{i}\mathbb{E}_{p_{\psi_i}(\mathbf{x}_i|\mathbf{s}_i)}\Big[H\left(Y_{i}\mid\mathbf{x}_{i}\right)\Big]-\lambda_{i}H_{_{\psi_{i}}}\left(Y_{i}\mid \mathbf{s}_{i}\right)\bigg\}.
	\end{aligned}
	\end{equation}
	Moreover,  because $p_{\psi_i}\left(\mathbf{x}_i\mid\mathbf{s}_i\right)=\prod_{r=1}^dp_{\psi_i}\left(x_{i,r}\mid\mathbf{s}_i\right)$ and the discrete memoryless   channel model $p(\mathbf{y}_i\mid\mathbf{x}_i)$,  $H_{\psi_i}\left(X_i\mid\mathbf{s}_i\right)$ and $H\left(Y_i\mid\mathbf{x}_i\right)$ can be respectively decomposed as
\begin{subequations}	
\begin{align}
&H_{\psi_{i}}\left(X_{i}\mid\mathbf{s}_{i}\right)=\sum_{r=1}^{d}H_{\psi_{i}}\left(X_{i,r}\mid\mathbf{s}_{i}\right), \\
	&H\big(Y_i\mid\mathbf{x}_i\big)=\sum_{r=1}^dH\big(Y_{i,r}\mid x_{i,r}\big).
	\end{align}
\end{subequations}
	Furthermore, by applying the reparameterization trick and Monte Carlo sampling, we   obtain an unbiased estimation of the gradient and thus optimize the objective using stochastic gradient descent.
	In particular, given a mini-batch of data $\left(\mathbf{s}_i^{(v)},\mathbf{u}_i^{(v)}\right)_{v=1}^V$ and sampling the
	channel noise $L$ times for each pair $\left(\mathbf{s}_i^{(v)},\mathbf{u}_i^{(v)}\right)$, we   obtain the Monte Carlo estimate     as follows
    {\small
		\begin{equation}\label{loss}
		\begin{aligned}
		 &\tilde{L}_{\mathrm{VRIB}}\left(\left\{\psi_{i},\theta_{i}\right\}_{i=1}^{N}\right)=\sum_{i=1}^{N}\frac{1}{V}\sum_{v=1}^{V}\biggr\{\frac{1}{L}\sum_{l=1}^{L}\Big[-\log q_{\theta_{i}}\left(\mathbf{u}_{i}^{(v)}\mid \right.\\
		 &\left.\mathbf{y}_{i}^{(v,l)}\right)+\lambda_{i}\sum_{r=1}^{d}H\left(Y_{i,r}\mid{x_{i,r}}^{(v,l)}\right)\Big]-\sum_{r=1}^{d}H_{\psi_{i}}\Big(X_{i,r}\mid{\mathbf{s}_{i}}^{(v)}\Big)\biggr\},
		\end{aligned}
		\end{equation}}where $\mathbf{y}_{i}{}^{(v,l)}=\left({y_{i,r}}^{(v,l)}\right)_{r=1}^{d}$ is the received signal, $x_{i,r}^{(v,l)}\sim p_{\psi_{i}}\left(x_{i,r}\mid\mathbf{s}_{i}^{(v)}\right)$ is the discrete feature representation,  ${{n}_{i,r}}^{(v,l)}\sim{\cal CN}\left(0,{\sigma_i}^{2}\right)$ is the channel noise, and
	${y_{i,r}}^{(v,l)}=g_{i,i}\Theta_{m}\left({x_{i,r}}^{(v,l)}\right)+g_{i,j}\sum_{j=1,j\neq i}^{N}\Theta_{m}\left({x_{i,r}}^{(v,l)}\right) +{{n}_{i,r}}^{(v,l)}.$

	The whole training process is summarized in Algorithm 1.
\begin{algorithm}[htb]
	\caption{Distributed SFDMA for Classification Tasks }
	\begin{algorithmic}[1]
		\State {\bf  Initialization:} Initializing  parameters $\left\{\psi_{i},\theta_{i}\right\}_{i=1}^{N}$;
		\State {\bf  Input:} Training data  $\left\{\mathbf{s}_i\right\}_{i=1}^{N}$,  transmitting power $p_i$,    number of epochs $T$, number of users $N$;
		\For { $t=1,..,T$}
		\For { $i=1,..,N$}
		\State \hspace*{0.2in}  Parameters of semantic encoder and decoder networks
		$\left\{ {{\psi _j},{\theta _j}} \right\}_{j = 1,j \ne i}^N \to \left\{ {{f_{{\psi _i}}}\left( . \right),{f_{{\theta _i}}}\left( . \right)} \right\}_{i = 1}^N$;
		\State{\bf TXs:}
		\State \hspace*{0.2in}
		$\left\{ {{f_{{\psi _i}}}\left( {{{\bf{s}}_i}} \right)} \right\}_{i = 1}^N \to \left\{ {{{\bf{x}}_i}} \right\}_{i = 1}^N$;
		\State \hspace*{0.2in}Transmit $\left\{\mathbf{x}_i\right\}_{i=1}^N$ over the channel;
		\State{\bf Channel:}
		\State \hspace*{0.2in}Randomly generate channel gains
		$\mathbf{g}_{i,j}\sim\mathcal{CN}\left(0,1\right),i,j\in\left\{1,..,N\right\}$;
		\State \hspace*{0.2in} Randomly generate AWGN $\left\{\mathbf{n}_i\right\}_{i=1}^N\sim\mathcal{CN}\left(0,{\sigma_i}^2\mathbf{I}\right)$;
		\State{\bf RXs:}
		\State \hspace*{0.2in}Receive  $\left\{\mathbf{y}_i\right\}_{i=1}^{N}$ by \eqref{y11};
		\State \hspace*{0.2in}Equalization  $\left\{\mathbf{\hat y}_i\right\}_{i=1}^{N}$ by \eqref{y1};
		\State \hspace*{0.2in}Semantic inference $\left\{f_{\theta_i}\left(\mathbf{\hat y}_i\right)\right\}_{i=1}^N\to\left\{\mathbf{\hat{u}}_i\right\}_{i=1}^N$;
		\State Fix network parameters 	$\left\{ {{\psi _j},{\theta _j}} \right\}_{j = 1,j \ne i}^N$ and optimize parameters	$\left\{ {{\psi _i},{\theta _i}} \right\}$ based on \eqref{loss};
		\State Update the parameters $\left\{ {{\psi _i},{\theta _i}} \right\}$  through backpropagation;
		\EndFor
		\EndFor
		\State{\bf Output:}  The parameters $\left\{\psi_{i},\theta_{i}\right\}_{i=1}^{N}$ of the semantic encoder and decoder networks
	\end{algorithmic}
\end{algorithm}

	\section{ Multi-user Semantic Interference  Network for Image Reconstruction }
	\begin{figure}[!t]
		\centering
		\includegraphics[width=14cm]{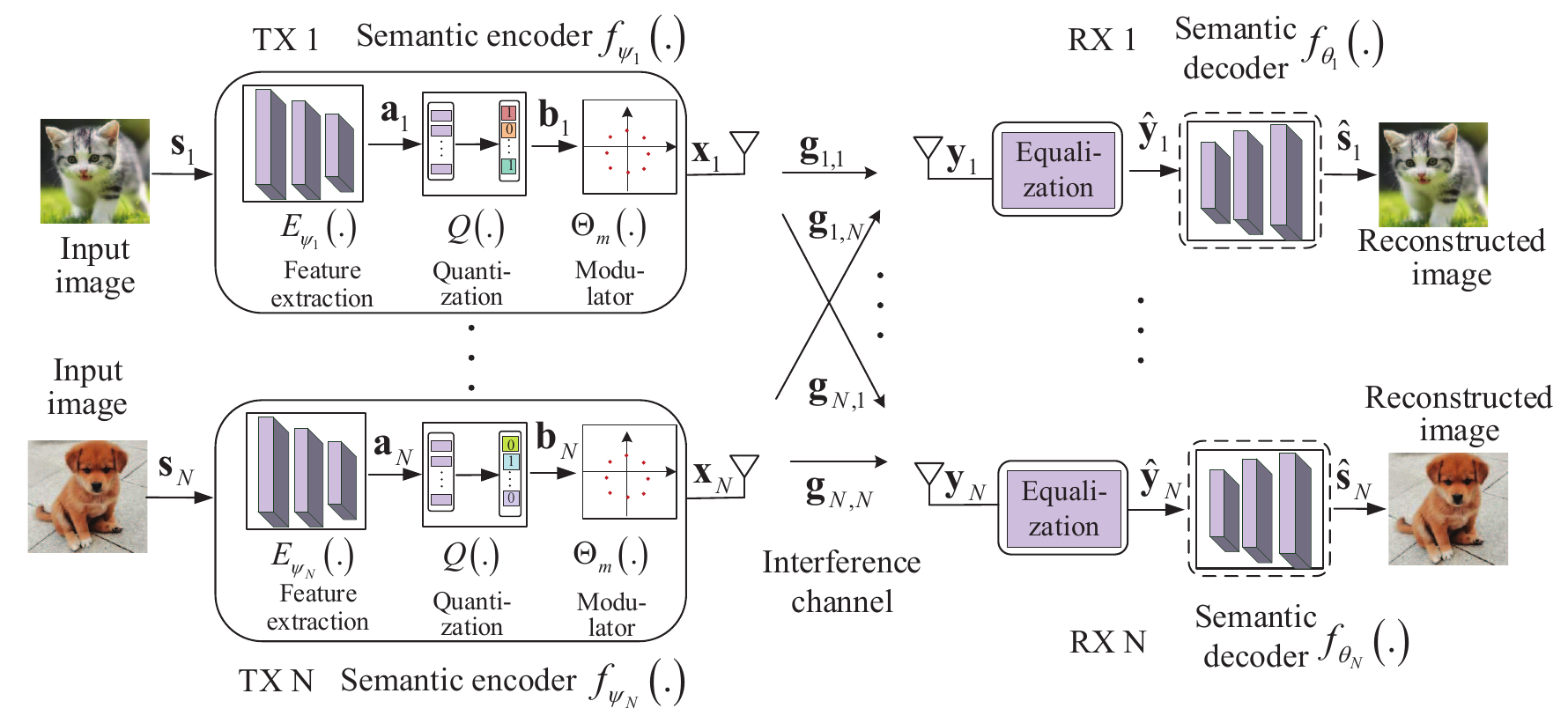}
		\caption{Multi-user semantic interference network for image reconstruction}
		\label{fig3}
	\end{figure}
	Note that, the inference task-oriented semantic encoders/decoders generally cannot be used for data reconstruction, and data-reconstruction semantic encoders/decoders are   more complex than inference task-oriented semantic encoders/decoders. Therefore, we further study semantic communication systems for data reconstruction.
	As shown in Fig. \ref{fig3}, the proposed semantic interference   network includes $N$      semantic encoder and decoder pairs, where $\mathbf{s}_i\in\mathbb{R}^{H\times W\times 3}$ denotes the input image of the semantic encoder, and  $i\in\{1,..,N\}$.
	
	The feature extraction network ${{\rm{E}}_{{\psi _i}}}\left( . \right)$ extracts and encodes  semantic feature $\mathbf{a}_i$ from $\mathbf{s}_i$, and then the semantic feature $\mathbf{a}_i$ is quantized to discrete representation $\mathbf{x}_i$. Note that, by jointly  training $N$ semantic encoder-decoder pairs, the semantic feature vector  $\mathbf{x}_i$ extracted by different users is approximately orthogonal, i.e.,
	${\bf{x}}_i^H{{\bf{x}}_j} \to 0,\forall j \ne i$.
	
\begin{figure}[!ht]
	\begin{minipage}[t]{0.5\textwidth}
		\centering
		\includegraphics[width=8cm]{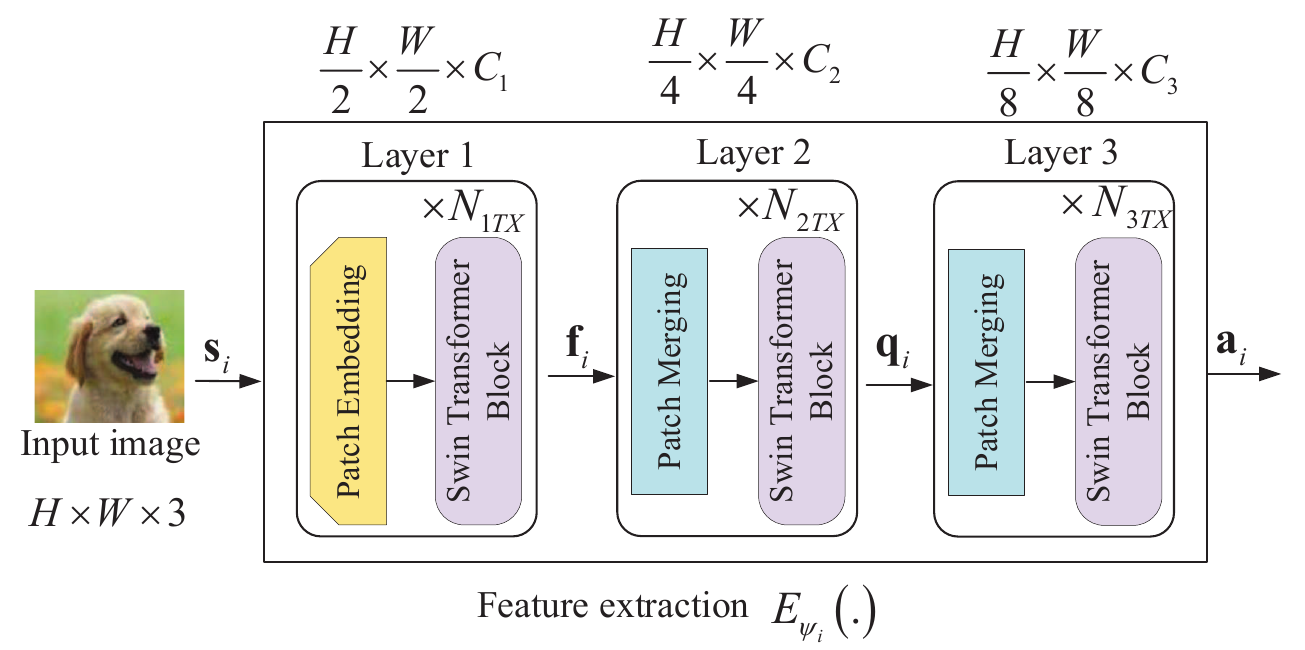}
		\vskip-0.2cm\centering {\footnotesize (a)}
	\end{minipage}
	\begin{minipage}[t]{0.5\textwidth}
		\centering
		\includegraphics[width=8cm]{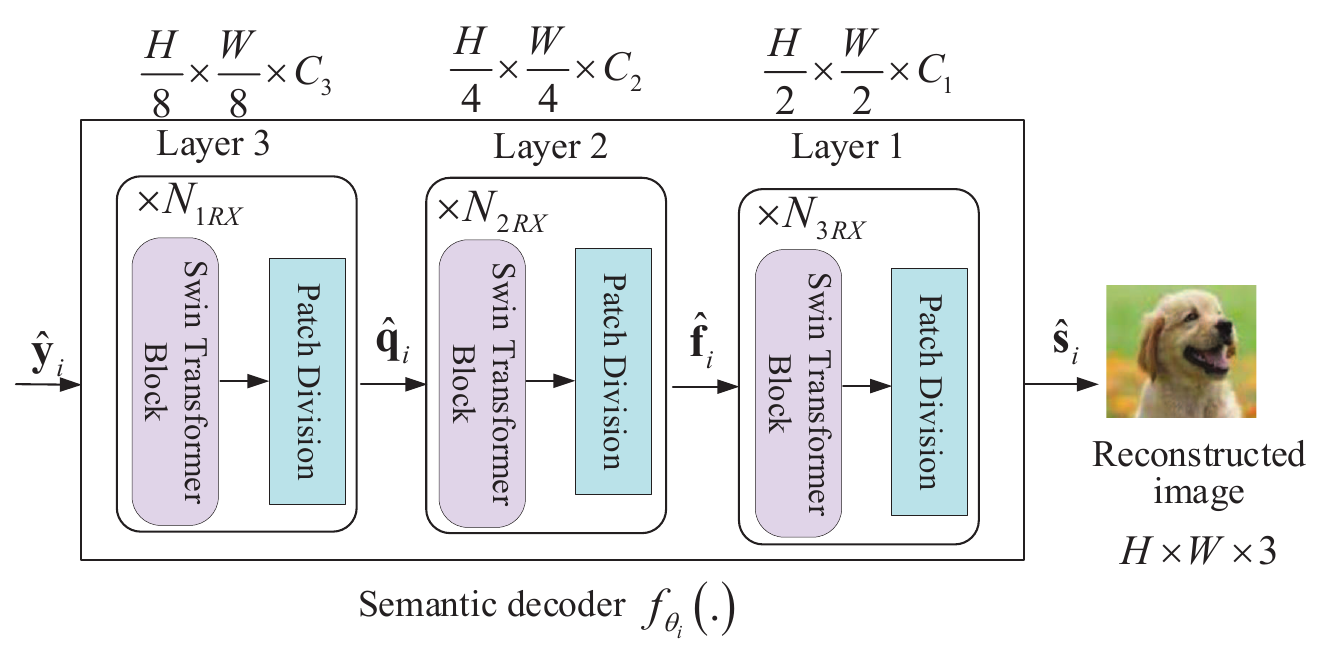}
		\vskip-0.2cm\centering {\footnotesize (b)}
	\end{minipage}
	\caption{ (a) The   feature extraction network. (b) The    semantic decoder network.}
	\label{fig4}
\end{figure}

	As shown in Fig. \ref{fig4} (a), the feature extraction  network $E_{\psi_i}\left(  \cdot  \right)$ of    TX  $i$ consists of three  layers. Specifically, in Layer $1$,  the input  image $\mathbf{s}_i$ is   divided    into $\frac{H}{2}\times\frac{W}{2}$ non-overlapping patches by a Patch Embedding layer $l_{\rm PE}(.)$, and then after patch embedding,  the non-overlapping   patches  are processed by $N_{_{\rm 1TX}}$ Swin Transformer Blocks $l_{\rm ST}(.)$ in sequence\cite{Liu_2021_ST}, i.e.,
	\begin{align}
	\mathbf{f}_{i}=l_{\rm ST_{N_{\rm 1TX}}}\Big(...l_{\rm ST_{1}}\Big(l_{\rm PE}\big(\mathbf{s}_{i}\big)\Big)\Big),
	\end{align}
	where  $\mathbf{f}_i\in\mathbb{R}^{\frac{H}{2}\times\frac{W}{2}\times C_1}$ denote the output of the $N_{_{1TX}}$ Swin Transformer Blocks.
	Note that, the Swin Transformer Block is a sequence-to-sequence function\cite{Yang_2023_WITT}, which consists of two sub-blocks. Each sub-block consists of a normalization layer, an attention module, followed by another normalization layer and an MLP layer. The first sub-block uses the Window MSA (W-MSA) module, while the second sub-block uses the Shifted Window MSA (SW-MSA) module.

	Furthermore, the $\mathbf{f}_i$ is fed to Layer 2 for downsampling through a Patch Merging layer $l_{\rm PM}(.)$, and the down-sampled data is   processed by $N_{_{\rm 2TX}}$ Swin Transformer Blocks $l_{\rm ST}(.)$, i.e.,
	\begin{align}
	\mathbf{q}_{i}=l_{\rm ST_{N_{\rm 2TX}}}\Big(...l_{\rm ST_{1}}\Big(l_{\rm PM}\big(\mathbf{f}_{i}\big)\Big)\Big),
	\end{align}
	where $\mathbf{q}_{i}\in\mathbb{R}^{\frac{H}{4}\times\frac{W}{4}\times C_2}$ denotes the output of Layer $2$.
	Moreover,    $\mathbf{q}_i$ is  processed by Layer 3, which includes a down-sampling Patch Merging layer $l_{\rm PM}(.)$ and $N_{_{\rm 3TX}}$ Swin Transformer Blocks $l_{\rm ST}(.)$. Finally,  the   extracted semantical  feature vector $\mathbf{a}_i\in\mathbb{R}^{\frac{H}{8}\times\frac{W}{8}\times C_3}$, i.e.,
	\begin{align}
	\mathbf{a}_{i}=l_{\rm ST_{N_{\rm 3TX}}}\Big(...l_{\rm ST_{1}}\Big(l_{\rm PM}\big(\mathbf{q}_{i}\big)\Big)\Big).
	\end{align}

	Then, using a quantizer $Q\left(  \cdot  \right)$, the feature vector $\mathbf{a}_i$ is   quantized into  bit streams $\mathbf{b}_i$, i.e.,
	\begin{align}
	\mathbf{b}_i=Q(\mathbf{a}_i),i\in\{1,..,N\}.
	\end{align}

	Then, the quantized  feature representations $\mathbf{b}_i$ is  digitally modulated and normalized into  $\mathbf{x}_{i}$, i.e.,

	\begin{align}
	\mathbf{x}_{i}={\rm{Norm}}\left(\Theta_{m}\left(\mathbf{b}_{i}\right)\right),i\in\{1,..,N\}.
	\end{align}
	
	In summary, in
	the  semantic encoder  $f_{\psi_{i}}\left(  \cdot  \right)$ includes semantic feature extraction, quantization and modulation, i.e.,
	\begin{align}
	\mathbf{x}_i=f_{\psi_i}\left(\mathbf{s}_i\right),i\in\left\{1,...,N\right\},
	\end{align}
	where $f_{\psi_i}\left(.\right)$ represents the semantic encoder of the TX  $i$,  and ${\psi_i}$ is the encoder parameter.

	Then, as shown in Fig. \ref{fig3},  the $N$ TXs simultaneously transmit    signals $\left\{ {{{\bf{x}}_i}} \right\}_{i = 1}^N$ to $N$ RXs,
	where  $\mathbf{g}_{i,i}$ is the channel gain from  the TX  $i$    to the RX  $i$,   $\mathbf{g}_{i,j}$ is the channel gain from TX $j$  to RX  $i$,  and  ${p_i}$ is the transmission power of the TX  $i$.
	The received signal of   the RX  $i$  $\mathbf{y}_{i},i\in\left\{1,...,N\right\}$ is  given by
	\begin{align}\label{y22}
	{{\bf{y}}_i} = {{\bf{g}}_{i,i}} \odot {\sqrt{p_i}} {{\bf{x}}_i} +  \sum\limits_{j = 1,j \ne i}^N {{\bf{g}}_{i,j}} \odot {\sqrt{p_j}} {{{\bf{x}}_j}}  + {{\bf{n}}_i},
	\end{align}
	where   ${{\bf{n}}_i}$ denotes the received additive white Gaussian noise of the RX  $i$, i.e., ${{\bf{n}}_i}\sim\mathcal{CN}\left(0,{\sigma_i}^2{\bf{I}}\right)$.
	
	 Furthermore,   assume that the channel state information is perfectly known, and therefore, the received signal ${{{\bf{\hat y}}}_i}$ after equalization \cite{Xie_VQA_2022}can be transformed to
	\begin{subequations}\label{y2}
		\begin{align}
		&{{{\bf{\hat y}}}_i} = {\left( {{\bf{g}}_{i,i}^H{{\bf{g}}_{i,i}}} \right)^{ -   1}}{\bf{g}}_{i,i}^H{{\bf{y}}_i}\\
		& = \sqrt {{p_i}} {{\bf{x}}_i} + \sum\limits_{j = 1,j  \ne i}^N {\bf{g}}_{i,i}^{ - 1}{{\bf{g}}_{i,j}} \odot {\sqrt {{p_j}} } {{\bf{x}}_j} + {\bf{g}}_{i,i}^{ - 1}{{\bf{n}}_i}.
		\end{align}
	\end{subequations}
	
	Finally, by exploiting  the  semantic  decoder $f_{\theta_{i}}\left(  \cdot  \right)$, the received signal $\mathbf{\hat y}_{i}$  is recovered as a  reconstructed image ${\hat {\bf{s}}_i}$.
	Specifically, the semantic decoder network  $f_{\theta_{i}}\left(  \cdot  \right)$, as shown in Fig. \ref{fig4} (b),     consists of  three layers.
	In detail, the feature vector  $\mathbf{y}_i\in\mathbb{R}^{\frac{H}{8}\times\frac{W}{8}\times C_3}$ is first upsampled by $N_{_{\rm 3RX}}$ Swin Transformer Blocks $l_{\rm ST}(.)$ and then by applying a Patch Division layer $l_{\rm PD}(.)$, we  obtain $\mathbf{\hat{q}}_i$ as
	\begin{align} \hat{\mathbf{q}}_{i}=l_{\rm PD}\Big(l_{\rm ST_{N_{\rm 3RX}}}\Big(...l_{\rm ST_{1}}\big(\mathbf{\hat y}_{i}\big)\Big)\Big).
	\end{align}
	Then, $\mathbf{\hat{q}}_i\in\mathbb{R}^{\frac{H}{4}\times\frac{W}{4}\times C_2}$ is fed to Layer 2, which includes $N_{_{\rm 2RX}}$ Swin Transformer Blocks $l_{\rm ST}(.)$ and an up-sampling Patch Division layer $l_{\rm PD}(.)$. The output of   Layer 2 is given by
	\begin{align}
	 \hat{\mathbf{f}}_{i}=l_{\rm PD}\Big(l_{\rm ST_{N_{\rm 2RX}}}\Big(...l_{\rm ST_{1}}\big(\hat{\mathbf{q}}_{i}\big)\Big)\Big).
	\end{align}
	Finally, $\mathbf{\hat{f}}_i\in\mathbb{R}^{\frac{H}{2}\times\frac{W}{2}\times C_1}$  is sent into $N_{_{\rm 1RX}}$ Swin Transformer Blocks $l_{\rm ST}(.)$ and an up-sampling Patch Division layer $l_{\rm PD}(.)$, and the   reconstructed image ${\hat {\bf{s}}_i}\in\mathbb{R}^{H\times W\times 3}$ is given as
	\begin{align}
	 \mathbf{\hat{s}}_{i}=l_{\rm PD}\Big(l_{\rm ST_{N_{\rm 1RX}}}\Big(...l_{\rm ST_{1}}\big(\mathbf{\hat{f}}_{i}\big)\Big)\Big).
	\end{align}
	
	In short,
	the  semantic decoding process can be formulated  as, i.e.,
	\begin{align}
	\mathbf{\hat{s}}_i=f_{\theta_i}\left(\mathbf{\hat y}_i\right),i\in\left\{1,...,N\right\},
	\end{align}
	where $f_{\theta_{i}}\left(  \cdot  \right)$ represents the semantic decoder of the RX  $i$,  and   ${\theta_i}$ denotes the  parameter set of the semantic  decoder network.
		Therefore, the training loss function of the semantic interference network is given as
	\begin{align}\label{loss2}
	\min_{\{\psi_i,\theta_i\}_{i=1}^N}\sum_{i=1}^N\mathbb{E}_{\mathbf{s}_i\sim p{\left(\mathbf{s}_i\right)}}\mathbb{E}_{\hat{\mathbf{s}}_i\sim p{\left(\hat{\mathbf{s}}_i|\mathbf{s}_i\right)}}\begin{bmatrix}\text{MSE}(\mathbf{s}_i,\hat{\mathbf{s}}_i)\end{bmatrix},
	\end{align}
	where $\text{MSE}\left(  \cdot  \right)$  computes the mean squared error between images.

%

	The whole training process is shown in Algorithm 2.
	\begin{algorithm}[htb]
	\caption{Distributed SFDMA for Image Reconstruction }
	\begin{algorithmic}[1]
		\State {\bf  Initialization:}  Load pre-trained model and initialize parameters $\left\{\psi_{i},\theta_{i}\right\}_{i=1}^{N}$;
		\State {\bf  Input:} Training data  $\left\{\mathbf{s}_i\right\}_{i=1}^{N}$,  number of epochs $T$, number of users $N$;
		\For { $t=1,..,T$}
		\For { $i=1,..,N$}
		\State \hspace*{0.2in}Parameters of the semantic encoder and decoder networks
		$\left\{ {{\psi _j},{\theta _j}} \right\}_{j = 1,j \ne i}^N \to \left\{ {{f_{{\psi _i}}}\left( . \right),{f_{{\theta _i}}}\left( . \right)} \right\}_{i = 1}^N$;
		\State{\bf TXs:}
		\State \hspace*{0.2in}
		$\left\{ {{f_{{\psi _i}}}\left( {{{\bf{s}}_i}} \right)} \right\}_{i = 1}^N \to \left\{ {{{\bf{x}}_i}} \right\}_{i = 1}^N$;
		\State \hspace*{0.2in}Transmit $\left\{\mathbf{x}_i\right\}_{i=1}^N$ over the channel;
		\State{\bf Channel:}
		\State \hspace*{0.2in}Randomly generate channel gains
		$\mathbf{g}_{i,j}\sim\mathcal{CN}\left(0,1\right),i,j\in\left\{1,..,N\right\}$;
		\State \hspace*{0.2in} Randomly generate AWGN $\left\{\mathbf{n}_i\right\}_{i=1}^N\sim\mathcal{CN}\left(0,{\sigma_i}^2\mathbf{I}\right)$;
		\State{\bf RXs:}
		\State \hspace*{0.2in}Receive  $\left\{\mathbf{y}_i\right\}_{i=1}^{N}$ by \eqref{y22};
		\State \hspace*{0.2in}Equalization  $\left\{\mathbf{\hat y}_i\right\}_{i=1}^{N}$ by \eqref{y2};
		\State \hspace*{0.2in}Reconstructed images $\left\{f_{\theta_i}\left(\mathbf{\hat y}_i\right)\right\}_{i=1}^N\to\left\{\mathbf{\hat{s}}_i\right\}_{i=1}^N$;
		\State Fix network parameters 	$\left\{ {{\psi _j},{\theta _j}} \right\}_{j = 1,j \ne i}^N$ and optimize parameters	$\left\{ {{\psi _i},{\theta _i}} \right\}$ based on \eqref{loss2};
		\State Update the parameters $\left\{ {{\psi _i},{\theta _i}} \right\}$  through backpropagation;
		\EndFor
		\EndFor
		\State{\bf Output:} The parameters $\left\{\psi_{i},\theta_{i}\right\}_{i=1}^{N}$ of the semantic encoder and decoder networks.
	\end{algorithmic}
\end{algorithm}



\section{ ABG Formula and   Adaptive Power Control }

So far, the relationship between  end-to-end performance measurements and  transmit power has not been established, and thus  the theoretical basis for  adaptive power control design for  semantic communications is unknown, which leads to performance degradation   in random fading channels.
  To address this challenge, we  analyzed a large number of experiments    results of semantic communication networks, and found that, as SNR increases,
  inference accuracy fist rapidly  increases, and   then slowly increases to the upper bound, and then remains unchanged, which  has also been verified in \cite{Huang_IoT_2024,Xie_2023,Hu_TWC_2023,Jankowski_2021,Shao_2022}.
  Inspired by this phenomenon,  we propose  ABG formula to  approximately fit   the relationship between inference accuracy   and transmission power.
 Specifically, for the semantic communication networks   with  inference tasks in section III, the   relationship between classification accuracy of the $i$th transmission pair ${\phi _i}$   and  ${p_i}$ can be approximated as   ABG formula ${\phi _i}$ as follows
  	\begin{align}{\phi _i} = {\alpha _i} - \frac{{{\gamma _i}}}{{1 + {{\left( {{\beta _i}\frac{{{p_i}{{\left| {{{\bf{g}}_{i,i}}} \right|}^2}}}{{\sum\limits_{j = 1,j \ne i}^N {{p_j}{{\left| {{{\bf{g}}_{i,j}}} \right|}^2}}  + \sigma _i^2}}} \right)}^{{\tau _i}}}}},
 	\end{align}
 where ${\alpha _i}$ ${{\beta _i}}$ ${{\gamma _i}}$ and ${{\tau _i}}$ are  parameters of the ABG formula. Given the DL based semantic encoders and decoders, the   parameters ${\alpha _i}$ ${{\beta _i}}$ ${{\gamma _i}}$ and ${{\tau _i}}$ can be obtained through testing.

Considering     practical time-varying random fading channels $\left\{ {{{\bf{g}}_{i,j}}\left( t \right)} \right\}$, the  classification accuracy threshold of  the $i$th transmission pair is   ${\eta _i}$, i.e., ${\phi _i} \ge {\eta _i}$. Hence, the optimal power control of the $i$th transmission pair $p_i^*\left( t \right)$ is given as
	\begin{align}p_i^*\left( t \right) = \frac{{\sum\limits_{j = 1,j \ne i}^N {{p_j}\left( t \right){{\left| {{{\bf{g}}_{i,j}}\left( t \right)} \right|}^2}}  + \sigma _i^2}}{{{\beta _i}{{\left| {{{\bf{g}}_{i,i}}\left( t \right)} \right|}^2}}}{\left( {\frac{{{\gamma _i}}}{{{\alpha _i} - {\eta _i}}} - 1} \right)^{\frac{1}{{{\tau _i}}}}},	\end{align}
where $i=1,...,N$.

\section{ Experimental Results and Analysis}
\subsection{ Experimental Settings}

In this section, we evaluate the performance of our proposed semantic SFMDA
schemes with binary phase shift keying (BPSK) modulation on MNIST dataset and CelebFaces Attributes (CelebA) dataset.
The MNIST dataset includes a training set of 60, 000 gray-scale images and a test set of 10, 000 sample images with handwritten character digits from 0 to 9. We use it for experiments on classification tasks.
The CelebA dataset  is a large-scale face image dataset containing 202,599 face images of 10,177 celebrity identities. We use it to experiment with image reconstruction. The input image size for the semantic encoder
is $28\times28\times1$ for MNIST and $64\times64\times3$ for  CelebA.

The proposed semantic SFMDA
schemes  are
implemented in Pytorch and trained with 3070Ti
GPU. We employ the Adam optimization framework for
back-propagation, which represents a variant of the stochastic
gradient descent. 
When training the semantic encoder and decoder,
we use the loss defined in (25) for image classification tasks
while using the MSE loss in (38) for image reconstruction
tasks. TABLE \ref{table2} shows   parameters settings for image reconstruction.
	\begin{table}[htbp]
	\caption{Network Parameters}
	\label{table2}
	\centering
	\scalebox{0.8}{
	\begin{tabular}{|c|c|c|c|}
		\hline
		\rule{0pt}{7pt}   Parameters of TXs &    Value& Parameters of RXs &    Value  \\ \hline
		
		\rule{0pt}{6.5pt}$N_{\rm 1TX}$ &  2 & $N_{\rm 1RX}$ &  2\\ \hline
		
		\rule{0pt}{6.5pt}$N_{\rm 2TX}$ & 4 & $N_{\rm 2RX}$ & 4 \\ \hline
		
		\rule{0pt}{6.5pt}$N_{\rm 3TX}$ & 4 & $N_{\rm 3RX}$ & 4 \\ \hline
		
		\rule{0pt}{6.5pt}$C_1 $ &  128 & $C_1 $ &  128 \\ \hline
		
		\rule{0pt}{6.5pt}$C_2 $ & 256 & $C_2 $ &  256 \\ \hline
		
		\rule{0pt}{6.5pt}$C_3 $ &  512 & $C_3$ &  512\\ \hline
	
	\end{tabular}
}
\end{table}

In this section, we compare the following JSCC schemes:
\begin{itemize}
	\item[$\bullet$] Deep JSCC: Based on Deep JSCC,  multi-user interference is ignored during   training  phase  for each transmission pair, while the multi-user interference is considered during testing phase.
	
		\item[$\bullet$] Upper bound: By applying Deep JSCC,  multi-user interference
  is ignored during both training and testing phases for each transmission pair.

	\item[$\bullet$] SFDMA: Based on the  proposed  SFDMA,  the transmission pairs are jointly trained with multi-user interference.
	
	\item[$\bullet$] Distributed SFDMA: Based on the  proposed distributed  SFDMA,  the transmission pairs are  trained in a distributed fashion with multi-user interference.

\end{itemize}

\subsection{ Experimental results and analysis  for classification task}

\subsubsection{Distinguishable feature domains}

\begin{figure}[!ht]
	\begin{minipage}[t]{0.5\textwidth}
		\centering
		\includegraphics[width=6cm]{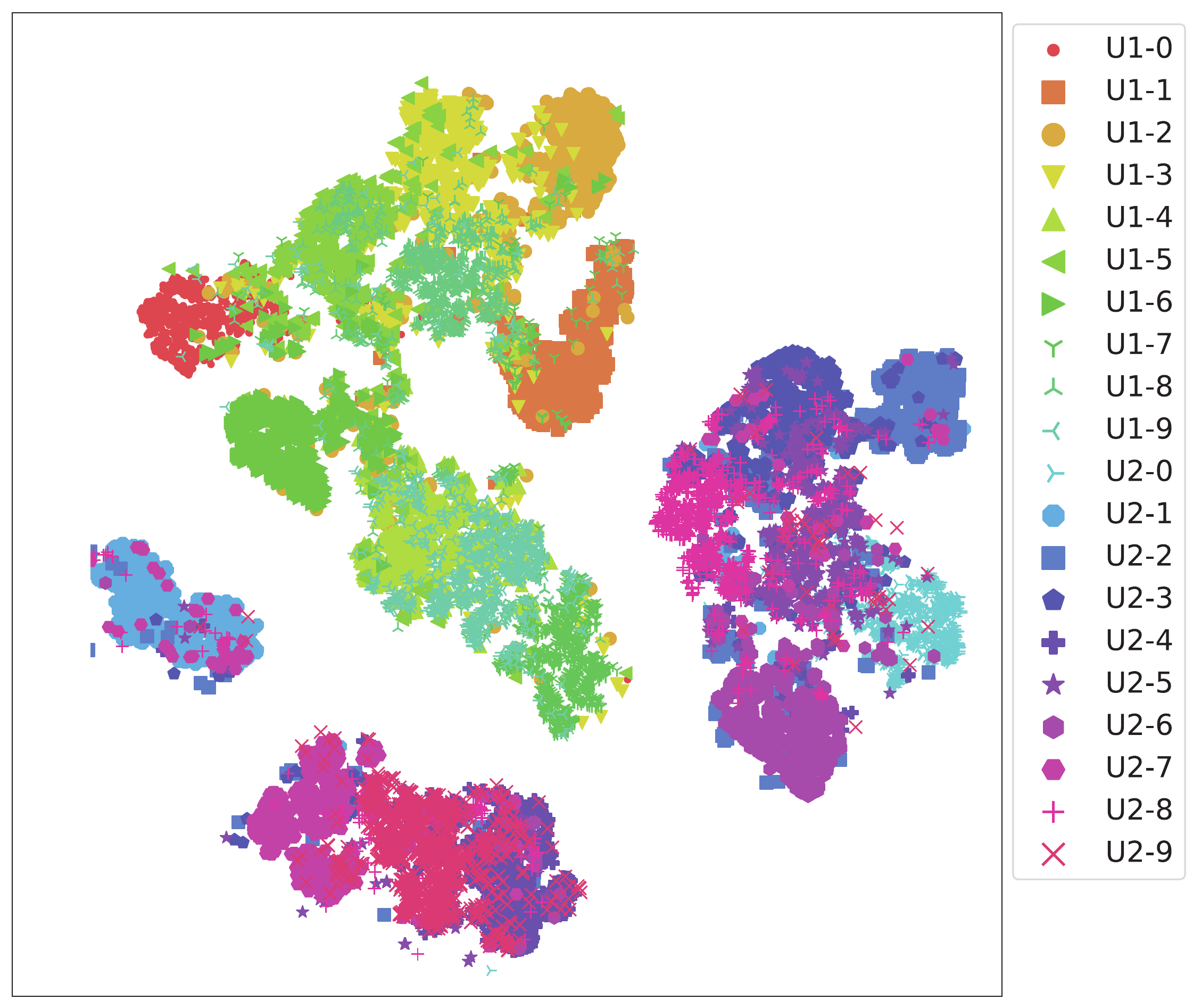}
		\vskip-0.2cm\centering {\footnotesize (a) $d=8$ bits,
			Acc1=69.73\%, Acc2=66.51\%.}
	\end{minipage}
	\begin{minipage}[t]{0.5\textwidth}
		\centering
		\includegraphics[width=6cm]{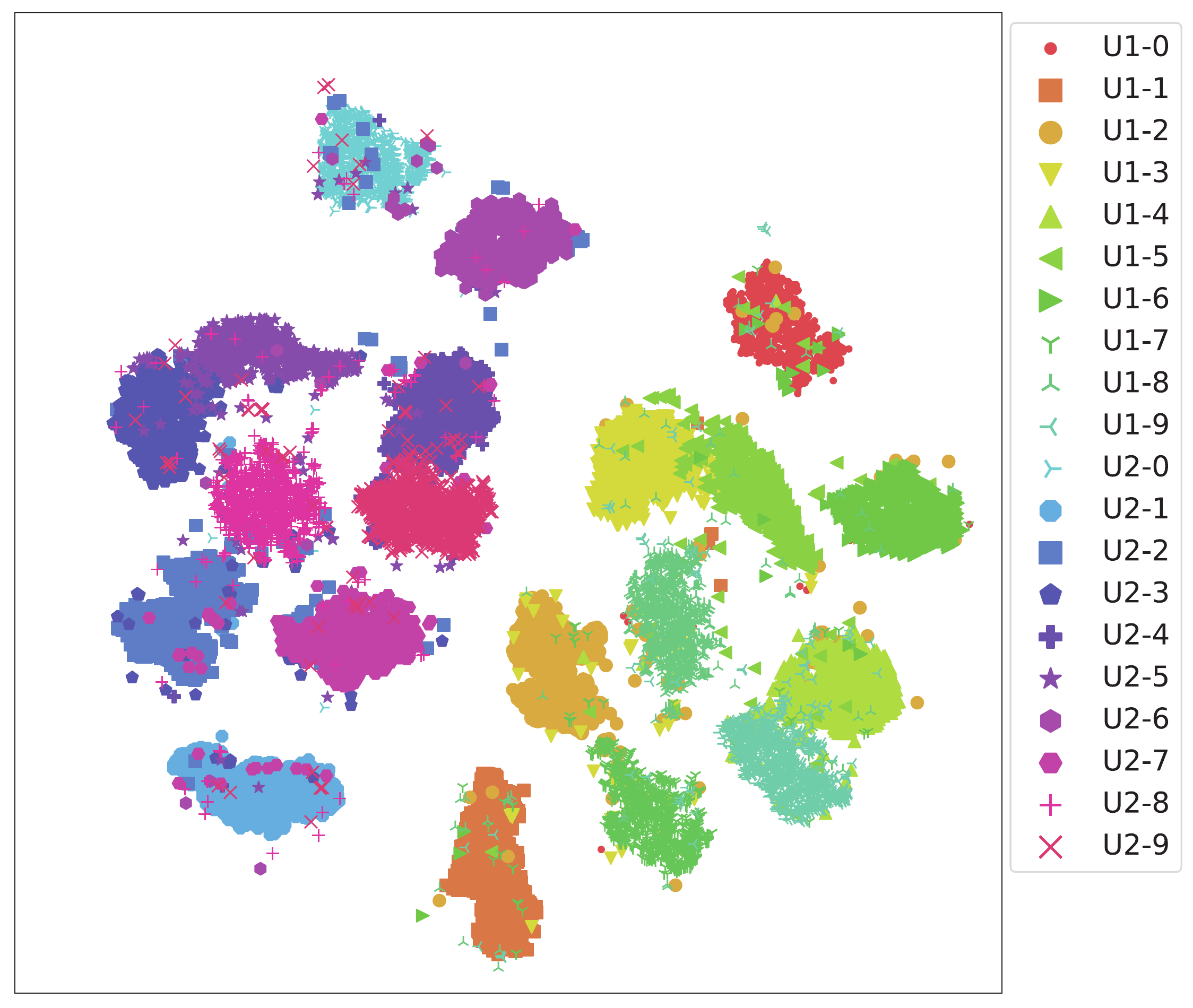}
		\vskip-0.2cm\centering {\footnotesize (b)  $d=64$ bits,
			Acc1=92.79\%, Acc2=92.93\%.}
	\end{minipage}
	\caption{  Distinguishable feature domain  of  TX $1$  and TX $2$  at  training SNR = 5dB.}
	\label{fig5}
\end{figure}

First, we evaluate   the distinguishability of the semantic feature domains of   the proposed SFMDA scheme  with different quantization bits.
In order to display the similarity of high-dimensional semantic feature vectors, we adopt the dimension reduction technique of t-SNE (t-Distributed Stochastic Neighbor Embedding)\cite{van2008visualizing}
to project
the semantic encoded signals  of the SFDMA network   into a 2-dimensional.  Fig. \ref{fig5} shows the 2-dimensional projections of the semantic encoded signals   $\mathbf{\hat x}_{1}$ and $\mathbf{\hat x}_{2}$ of SFDMA
  for       the  MNIST dataset, and  U1-$k$ represents the label $k$ of  $\mathbf{\hat x}_{1}$, and    U2-$l$ represents the label $l$ of  $\mathbf{\hat x}_{2}$, where $k,l \in \left\{ {{\rm{0}},{\rm{1}},...,9} \right\}$ are labels of the MNIST dataset.
Fig. \ref{fig5} (a) and  (b)  show    the distinguishability of the semantic feature domains of  SFDMA scheme with   $d=8$ bits and $d=64$ bits quantization, respectively,  where   training SNR=5dB.
In Fig. \ref{fig5} (a),  the classification accuracies of the 1st and 2nd   transmission pairs are $69.73\%$ and $66.51\%$ respectively, while In Fig. \ref{fig5} (b), the the classification accuracies of the 1st and 2nd   transmission pairs are $92.79\%$, $92.93\%$, respectively,
 Moreover, as the number of quantization bits increases, the distinction between different semantic features becomes   clearer, and the classification accuracies of the two semantic receivers increase.

\subsubsection{Orthogonality of  semantic features}

Furthermore, we investigate the orthogonality of the   semantic features  of     the  SFDMA   scheme.

\begin{table}[!ht]
	\caption{The   classification accuracy of  the proposed SFDMA scheme of two  transmission pairs  with the same inputs.}
	\centering
	\scalebox{0.8}{
		\begin{tabular}{|c|c|c|c|c|}
			\hline
			\rule{0pt}{15pt}	&$f_{\theta_{1}}(\mathbf{y}_{1})$ &$f_{\theta_1}\left(\mathbf{g}_{1,2}\mathbf{x}_2\right)$ & $f_{\theta_{2}}(\mathbf{y}_{2})$ & $f_{\theta_{2}}\left(\mathbf{g}_{2,1}\mathbf{x}_{1}\right)$ \\ \hline
			\rule{0pt}{15pt}  Accuracy(\%)  &92.37 & 9.71 & 92.19 & 10.01 \\ \hline	
		\end{tabular}
		\label{table3}
	}
\end{table}

   Table \ref{table3} shows the classification accuracy    of  the proposed SFDMA scheme, where the   inputs of two  transmission pairs are the same. In   Table \ref{table3},   $f_{\theta_{1}}(\mathbf{y}_{1})$ denotes  RX $1$ decoding the received signal $\mathbf{y}_{1}$, and the corresponding classification accuracy is  $92.37\%$, while $f_{\theta_1}\left(\mathbf{g}_{1,2}\mathbf{x}_2\right)$ denotes    RX $1$ decoding the received interference  $\mathbf{x}_2$ from  TX $2$, and the corresponding classification accuracy is  $9.71\%$, which verifies the   approximate  orthogonality  between  semantic features $\mathbf{x}_1$ and $\mathbf{x}_2$. Moreover,
    $f_{\theta_{2}}(\mathbf{y}_{2})$ denotes    RX $2$ decoding the received signal $\mathbf{y}_{2}$, and the corresponding classification accuracy is  $92.19\%$, $f_{\theta_2}\left(\mathbf{g}_{2,1}\mathbf{x}_1\right)$ denotes    RX $2$ decoding the received interference  $\mathbf{x}_1$ from  TX $1$, and the corresponding classification accuracy is  $10.01\%$, which also verifies the   approximate orthogonality   between  semantic features $\mathbf{x}_1$ and $\mathbf{x}_2$.

\begin{table}[!ht]
	\caption{ The inner product and angle among the semantic features of  three  transmission pairs  with the same inputs.}
	\centering
	\scalebox{0.8}{
		\begin{tabular}{|c|c|c|c|}
			\hline
			\rule{0pt}{15pt} Inner product& ${\mathbf{x}}_1^H{{\mathbf{x}}_2}$& ${\mathbf{x}}_1^H{{\mathbf{x}}_3}$ & ${\mathbf{x}}_2^H{{\mathbf{x}}_3}$ \\ \hline
			
			\rule{0pt}{15pt}  Value&$0.0021$ &$0.0029$&$ 0.0035$ \\ \hline
			\rule{0pt}{15pt}   Angle& $\arccos \left( {\mathbf{x}}_1^H{{\mathbf{x}}_2} \right)$& $\arccos \left( {\mathbf{x}}_1^H{{\mathbf{x}}_3} \right)$ & $\arccos \left( {\mathbf{x}}_2^H{{\mathbf{x}}_3} \right)$\\ \hline
			
			\rule{0pt}{15pt}  Value$\left( {^ \circ } \right)$&$89.879$ &$89.833$&$ 89.799$ \\ \hline	 
		\end{tabular}
		\label{table4}
	}
\end{table}

Table \ref{table4} presents   the inner product and angle  among the semantic features ${{\bf{x}}_1}$,  ${{\bf{x}}_2}$ and ${{\bf{x}}_3}$, where ${{\bf{x}}_1}$, ${{\bf{x}}_2}$ and ${{\bf{x}}_3}$ are the semantic features of TX $1$ , TX $2$ and TX $3$
respectively.
Table \ref{fig4} shows that the  inner product of  ${{\bf{x}}_1}$ and ${{\bf{x}}_2}$ is $0.0021$, and the  angle between ${{\bf{x}}_1}$ and ${{\bf{x}}_2}$ is $89.879$ degrees. The  inner product of  ${{\bf{x}}_1}$ and ${{\bf{x}}_3}$ is $0.0029$, and the  angle between ${{\bf{x}}_1}$ and ${{\bf{x}}_3}$ is $89.833$ degrees. The  inner product of  ${{\bf{x}}_2}$ and ${{\bf{x}}_3}$ is $0.0035$, and the  angle between ${{\bf{x}}_2}$ and ${{\bf{x}}_3}$ is $89.799$ degrees. The feature vector of TX $1$ , the feature vector of TX $2$  and the feature vector of TX $3$  are approximately orthogonal, which verifies the separation of TXs' semantic features of  the  SFDMA scheme.

\subsubsection{Classification results and analysis }

%
%
%
%

\begin{figure}[!ht]
	\begin{minipage}[t]{0.5\textwidth}
		\centering
		\includegraphics[width=7cm]{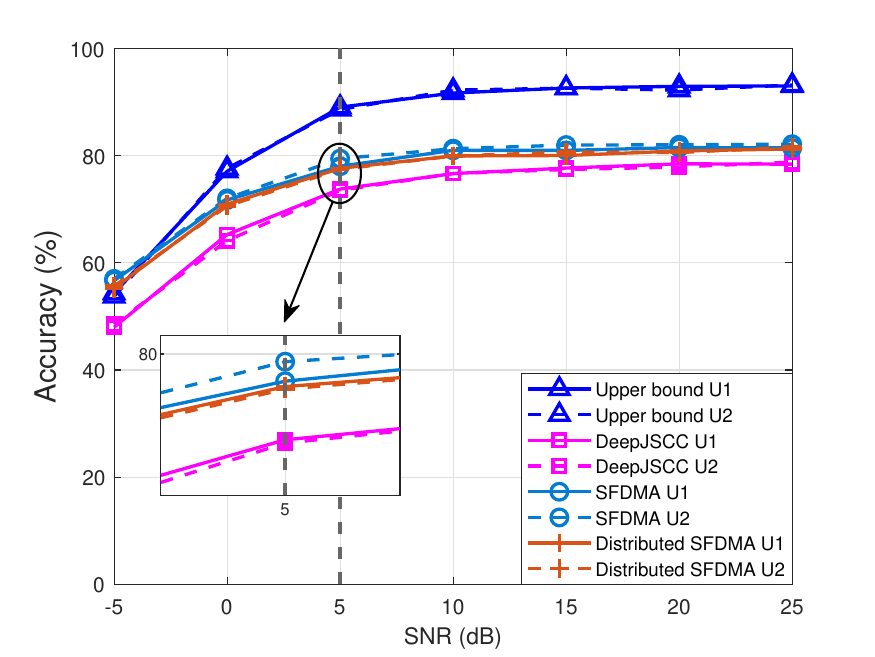}
		\vskip-0.1cm\centering {\footnotesize (a)  Training SNR=5dB, $d=64$ bits.}
	\end{minipage}
	\begin{minipage}[t]{0.5\textwidth}
		\centering
		\includegraphics[width=7cm]{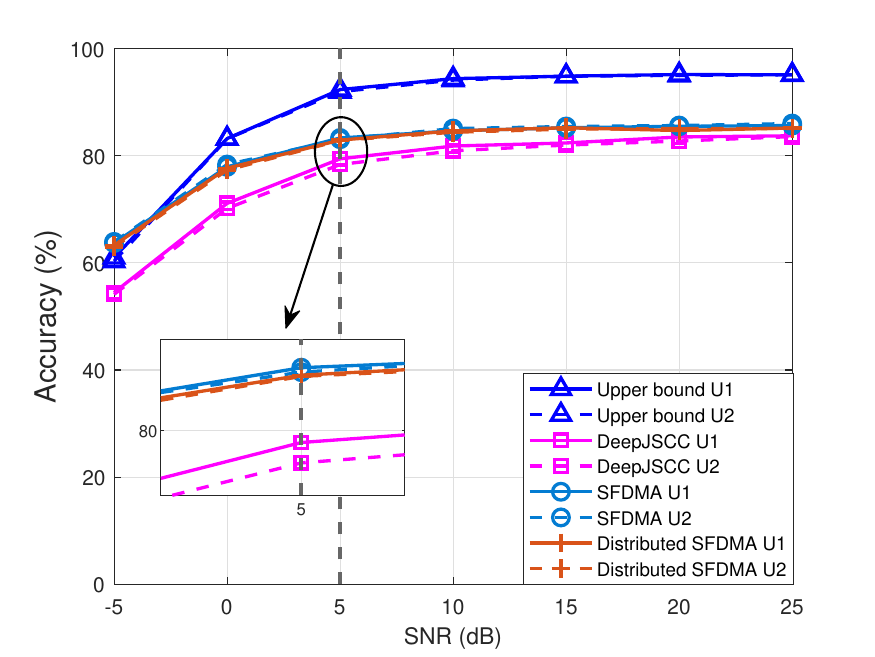}
		\vskip-0.1cm\centering {\footnotesize (b)  Training SNR=5dB, $d=128$ bits}
	\end{minipage}
	\caption{Performance of different   schemes   for classification task on the  MNIST  dataset  over Rayleigh channel with SNR  in [-5dB,25dB].  }
	\label{fig6}
\end{figure}

Furthermore, we compare the classification accuracy of  Upper bound,  Deep JSCC, SFDMA and distributed SFDMA schemes under different quantization bits on   Rayleigh channels.


Fig. \ref{fig6}  shows that the SFDMA scheme  on Rayleigh channel achieves a higher
classification accuracy than the existing deep JSCC approach with interference for all SNRs, especially at low SNR, the advantage of SFDMA scheme is more prominent. The classification accuracy of the  SFDMA scheme is lower than the  deep JSCC without interference in the medium and high-SNR regimes. However, in the low-SNR regime, the classification accuracy of the  SFDMA scheme is higher than the  deep JSCC without interference, because the SFDMA scheme is trained with interference and the  Deep JSCC is trained  without interference. Thus, the SFDMA scheme is more robust in the low-SNR regime than deep JSCC without interference.

This result validates the effectiveness and robustness of the proposed SFDMA scheme
in Sec. III.

\subsection{ Experimental results of image reconstruction}

To analyze the experimental results of image reconstruction, we adopt two evaluation metrics for the quality of image reconstruction peak signal-to-noise ratio (PSNR)  and
multi-scale structural similarity index metric (MS-SSIM)\cite{Zhu_2023_VIT}.

\subsubsection{Distinguishable feature domains}
%
%

\begin{figure}[h]
	\centering
	\includegraphics[width=7cm]{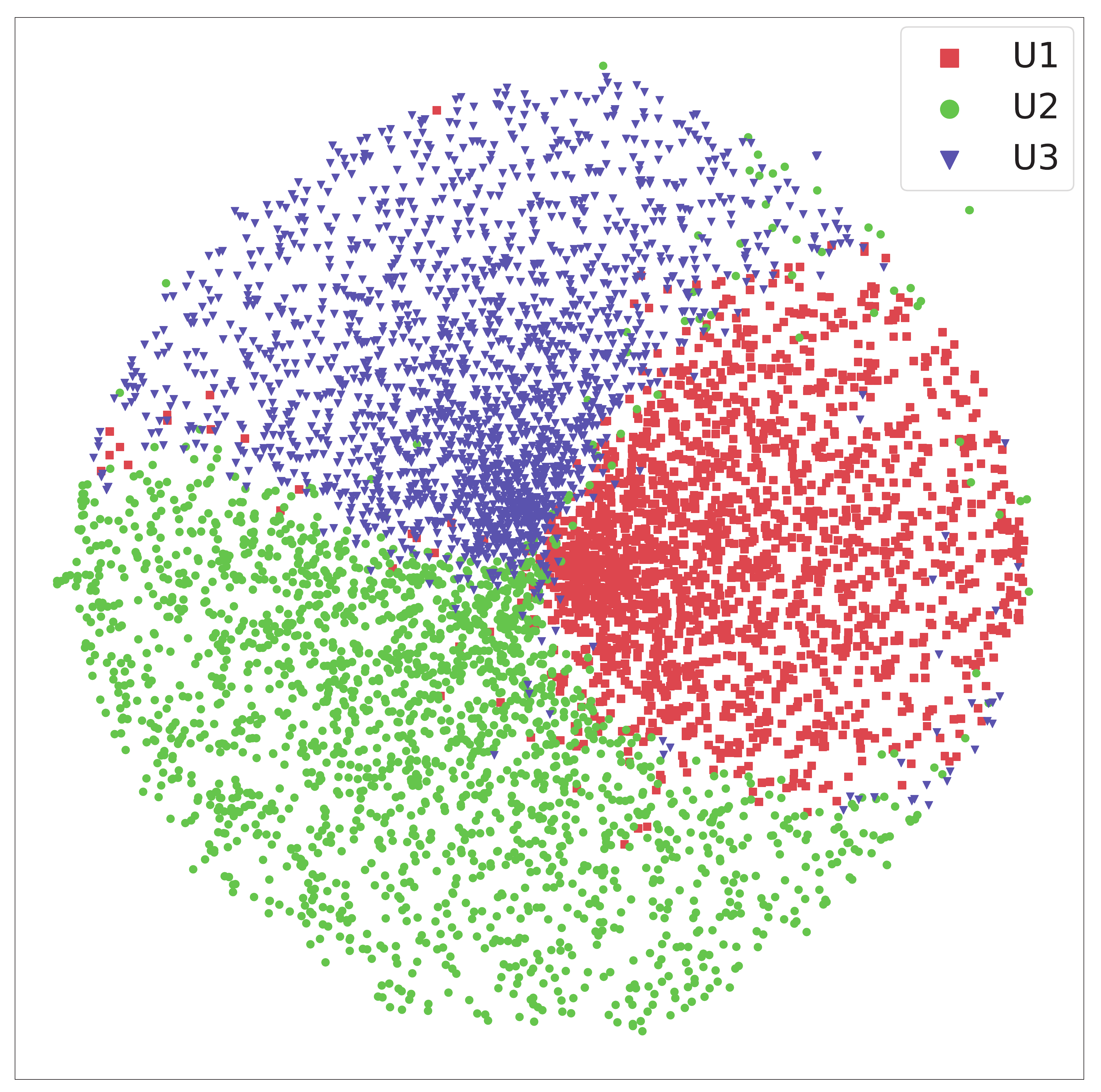}
	\caption{ Distinguishable feature domain  of  TX $1$, TX $2$ and  TX $3$}
	\label{fig7}
\end{figure}

 Fig. \ref{fig7} uses t-SNE to realize the visualization of semantic feature
dimensions, where U1, U2 and U3 represent the semantic features of TX $1$ $2$ and $3$, respectively.
Fig. \ref{fig7}   demonstrate sthe semantic feature discriminative performance of  three TXs under training SNR=5dB.
As illustrated in Fig. \ref{fig7},  the semantic features among different TXs are approximately distinguishable.


\subsubsection{Orthogonality of   semantic features}

\begin{table}[!ht]
	\caption{The    reconstructed images of  the proposed SFDMA scheme of two  transmission pairs  with the same inputs.}
	\centering
	\scalebox{0.8}{
		\begin{tabular}{|c|c|c|c|c|}
			\hline
			\rule{0pt}{15pt}	&$f_{\theta_{1}}(\mathbf{y}_{1})$ &$f_{\theta_1}\left(\mathbf{g}_{1,2}\mathbf{x}_2\right)$ & $f_{\theta_{2}}(\mathbf{y}_{2})$ & $f_{\theta_{2}}\left(\mathbf{g}_{2,1}\mathbf{x}_{1}\right)$ \\ \hline
			\rule{0pt}{15pt}  	PSNR(dB) & 25.647 &7.350& 25.675  &7.469 \\ \hline
			\rule{0pt}{15pt}  	MS-SSIM &0.928&0.066& 0.928  &0.086\\ \hline
		\end{tabular}
		\label{table5}
	}
\end{table}

 Table \ref{table5} shows reconstructed images of  the proposed SFDMA scheme of two  transmission pairs  with the same inputs. In   Table \ref{table5},   $f_{\theta_{1}}(\mathbf{y}_{1})$ denotes \ RX $1$ decoding the received signal $\mathbf{y}_{1}$, and the corresponding PSNR and MS-SSIM are $25.647$dB and $0.928$ respectively, while $f_{\theta_1}\left(\mathbf{g}_{1,2}\mathbf{x}_2\right)$ denotes    RX $1$ decoding the received interference  $\mathbf{x}_2$ from  TX $2$, and the corresponding PSNR and MS-SSIM are $7.350$dB and $0.066$ respectively, which verifies the   orthogonality  between  semantic features $\mathbf{x}_1$ and $\mathbf{x}_2$. Moreover,
    $f_{\theta_{2}}(\mathbf{y}_{2})$ denotes    RX $2$ decoding the received signal $\mathbf{y}_{2}$, and the corresponding PSNR and MS-SSIM are $25.675$dB and $0.928$ respectively, $f_{\theta_2}\left(\mathbf{g}_{2,1}\mathbf{x}_1\right)$ denotes    RX$2$ decoding the received interference  $\mathbf{x}_1$ from the  TX $1$, and the corresponding PSNR and MS-SSIM are $7.469$dB and $0.086$ respectively,  which also verifies  the approximately   orthogonality   between  semantic features $\mathbf{x}_1$ and $\mathbf{x}_2$.

Moreover, we investigate the orthogonality of JSCC semantic features  of TXs in the  SFDMA IC scheme.
Table \ref{table6} presents    the inner product and angle  among the semantic features ${{\bf{x}}_1}$,  ${{\bf{x}}_2}$ and ${{\bf{x}}_3}$, where ${{\bf{x}}_1}$, ${{\bf{x}}_2}$ and ${{\bf{x}}_3}$ are the semantic features of   TX $1$, TX $2$ and TX $3$,
respectively.

Table \ref{table6} shows that the  inner product of  ${{\bf{x}}_1}$ and ${{\bf{x}}_2}$ is $-0.0006$, and the  angle between ${{\bf{x}}_1}$ and ${{\bf{x}}_2}$ is $90.034$ degrees. The  inner product of  ${{\bf{x}}_1}$ and ${{\bf{x}}_3}$ is $0.0006$, and the  angle  between ${{\bf{x}}_1}$ and ${{\bf{x}}_3}$ is $89.965$ degrees. The  inner product of  ${{\bf{x}}_2}$ and ${{\bf{x}}_3}$ is $0.0031$, and the  angle between ${{\bf{x}}_2}$ and ${{\bf{x}}_3}$ is $90.177$ degrees. The feature vectors of TX $1$,  TX $2$  and  TX $3$  are approximately orthogonal, which verifies the separation of TXs' semantic features of  the  SFDMA scheme.

\begin{table}[!ht]
	\caption{ The inner product and angle among the semantic features of  three  transmission pairs  with the same inputs.}
	\centering
	\scalebox{0.8}{
		\begin{tabular}{|c|c|c|c|}
			\hline
			\rule{0pt}{15pt} Inner product& ${\mathbf{x}}_1^H{{\mathbf{x}}_2}$& ${\mathbf{x}}_1^H{{\mathbf{x}}_3}$ & ${\mathbf{x}}_2^H{{\mathbf{x}}_3}$ \\ \hline
			
			\rule{0pt}{15pt}  Value&$-0.0006$ &$0.0006$&$ -0.0031$ \\ \hline
			\rule{0pt}{15pt}   Angle& $\arccos \left( {\mathbf{x}}_1^H{{\mathbf{x}}_2} \right)$& $\arccos \left( {\mathbf{x}}_1^H{{\mathbf{x}}_3} \right)$ & $\arccos \left( {\mathbf{x}}_2^H{{\mathbf{x}}_3} \right)$\\ \hline
			
			\rule{0pt}{15pt}  Value$\left( {^ \circ } \right)$&$90.034$ &$89.965$&$ 90.177$ \\ \hline	 
		\end{tabular}
		\label{table6}
	}
\end{table}

\begin{figure*}
	\centering
	\subfigure{
		\begin{minipage}[t]{0.32\linewidth}
			\centering
			\includegraphics[width=5cm]{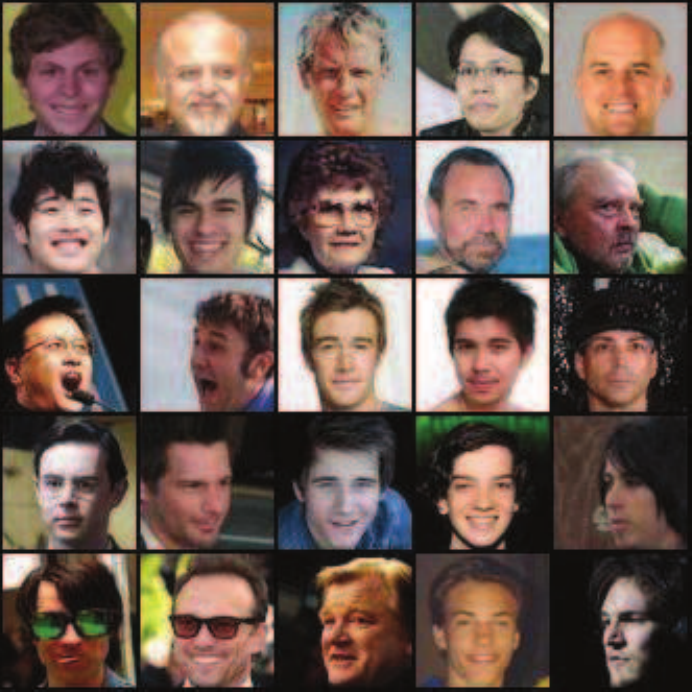}
			\vskip-0.1cm\centering {(a) Original image.}
		\end{minipage}
		\begin{minipage}[t]{0.32\linewidth}
		\centering
		\includegraphics[width=5cm]{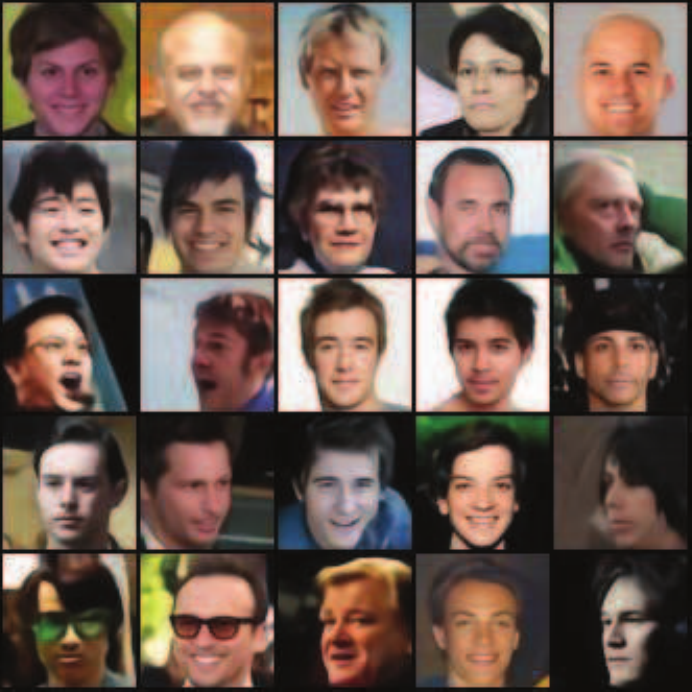}
		\vskip-0.1cm\centering {(b) Reconstructed image of  $f_{\theta_{1}}(\mathbf{y}_{1})$.}
    	\end{minipage}
    	\begin{minipage}[t]{0.32\linewidth}
    	\centering
    	\includegraphics[width=5cm]{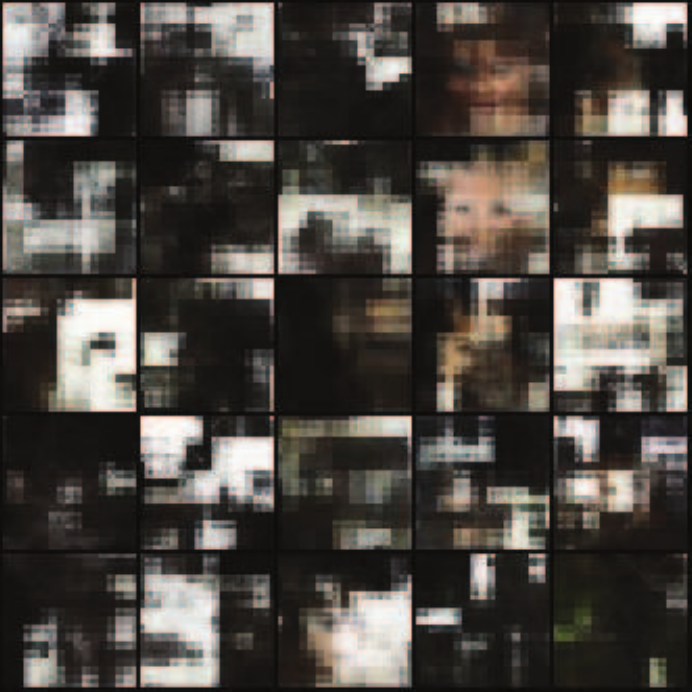}
    	\vskip-0.1cm\centering {(c) Reconstructed image of $f_{\theta_1}\left(\mathbf{g}_{1,2}\mathbf{x}_2\right)$.}
    \end{minipage}
}%

	\subfigure{
		\begin{minipage}[t]{0.5\linewidth}
			\centering
			\includegraphics[width=5cm]{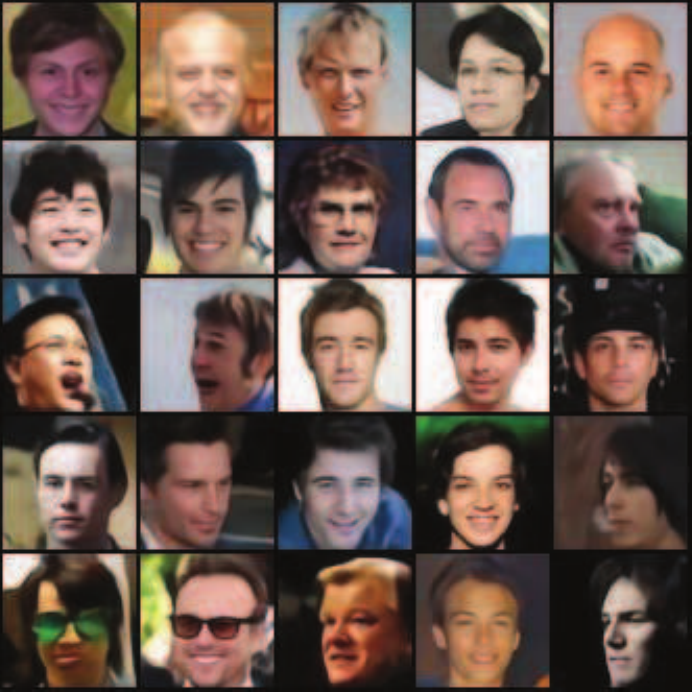}
			\vskip-0.1cm\centering {(d) Reconstructed image of $f_{\theta_{2}}(\mathbf{y}_{2})$.}
		\end{minipage}
	    \begin{minipage}[t]{0.5\linewidth}
		\centering
		\includegraphics[width=5cm]{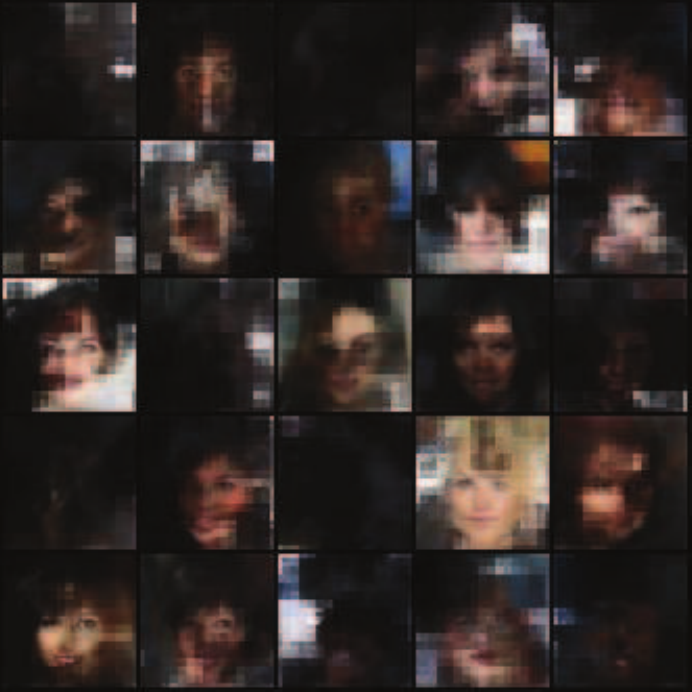}
		\vskip-0.1cm\centering {(e) Reconstructed image of $f_{\theta_{2}}\left(\mathbf{g}_{2,1}\mathbf{x}_{1}\right)$.}
     	\end{minipage}
	}%

	\centering
	\caption{Reconstructed images with different information decoded by different users. (a) Original image, (b) Reconstructed image of  $f_{\theta_{1}}(\mathbf{y}_{1})$, (c) Reconstructed image of $f_{\theta_1}\left(\mathbf{g}_{1,2}\mathbf{x}_2\right)$, (d) Reconstructed image of $f_{\theta_{2}}(\mathbf{y}_{2})$ and (e) Reconstructed image of $f_{\theta_{2}}\left(\mathbf{g}_{2,1}\mathbf{x}_{1}\right)$.   }
	\label{fig8}
\end{figure*}

\subsubsection{Reconstruction results and analysis}


\begin{figure}[!ht]
	\begin{minipage}[t]{0.5\textwidth}
		\centering
		\includegraphics[width=7.5cm]{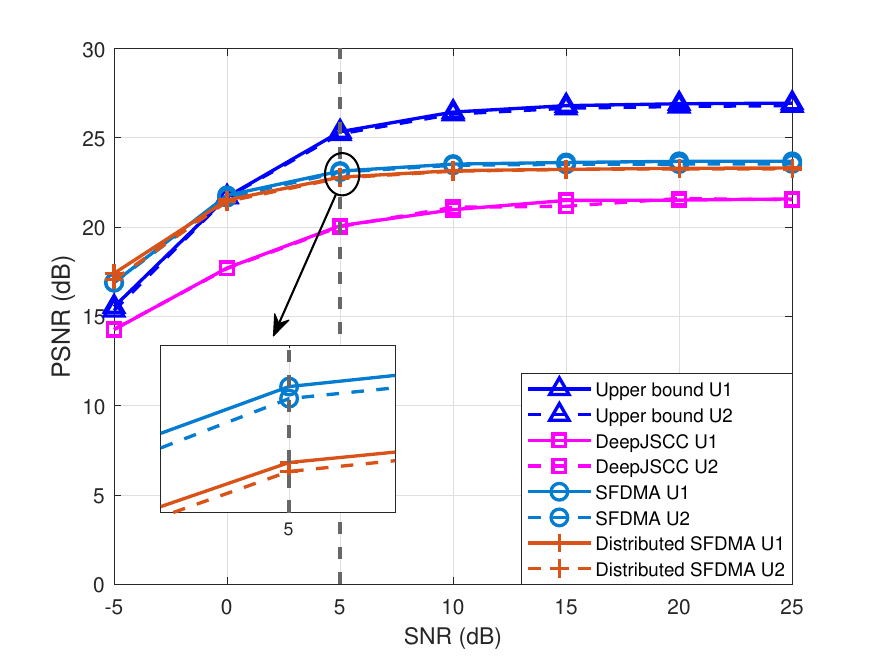}
		\vskip-0.1cm\centering {\footnotesize (a)  Training SNR=5dB, PSNR evaluation.}
	\end{minipage}
	\begin{minipage}[t]{0.5\textwidth}
		\centering
		\includegraphics[width=7.5cm]{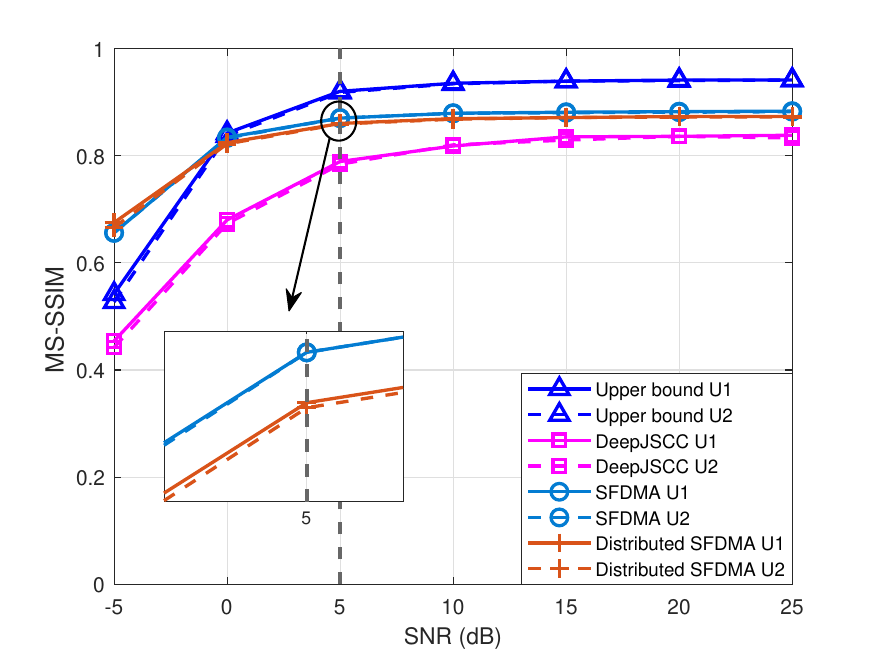}
		\vskip-0.1cm\centering {\footnotesize (b)  Training SNR=5dB, MS-SSIM evaluation.}
	\end{minipage}
	\caption{ Performance of different   schemes   for image reconstruction over   Rayleigh channel with SNR  in [-5dB,25dB].}
	\label{fig9}
\end{figure}

\begin{figure*}[!ht]
	\begin{minipage}[t]{0.32\textwidth}
		\centering
		\includegraphics[width=5cm]{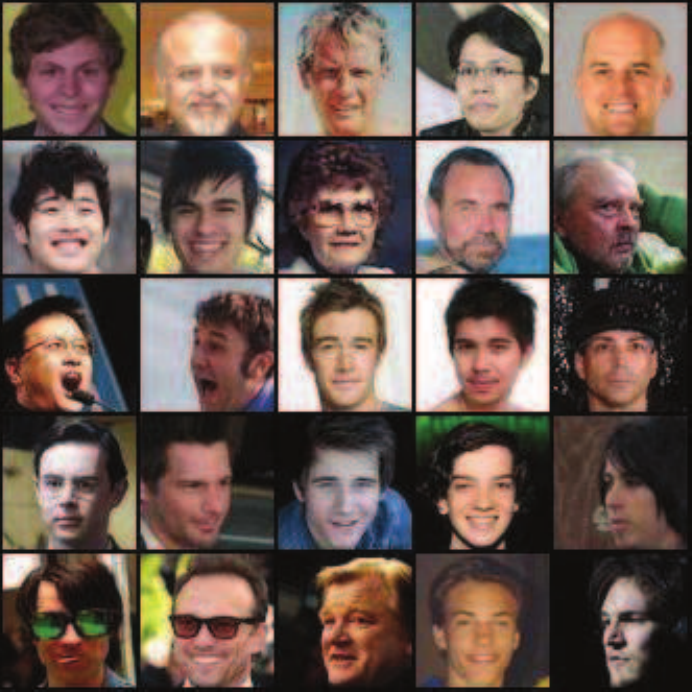}
		\vskip-0.1cm\centering {\footnotesize (a) Original image}
	\end{minipage}
	\begin{minipage}[t]{0.32\textwidth}
		\centering
		\includegraphics[width=5cm]{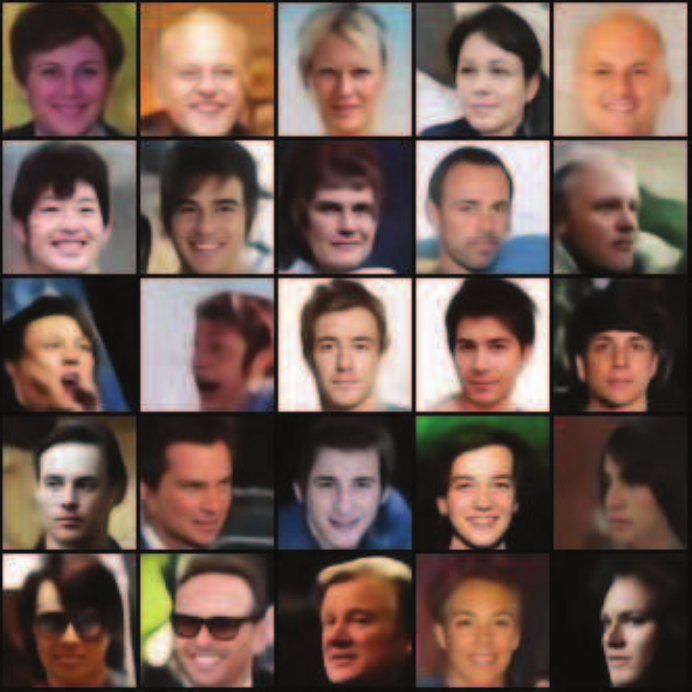}
		\vskip-0.1cm\centering {\footnotesize (b)  Reconstructed image. Training SNR=0dB}
	\end{minipage}
	\begin{minipage}[t]{0.32\textwidth}
		\centering
		\includegraphics[width=5cm]{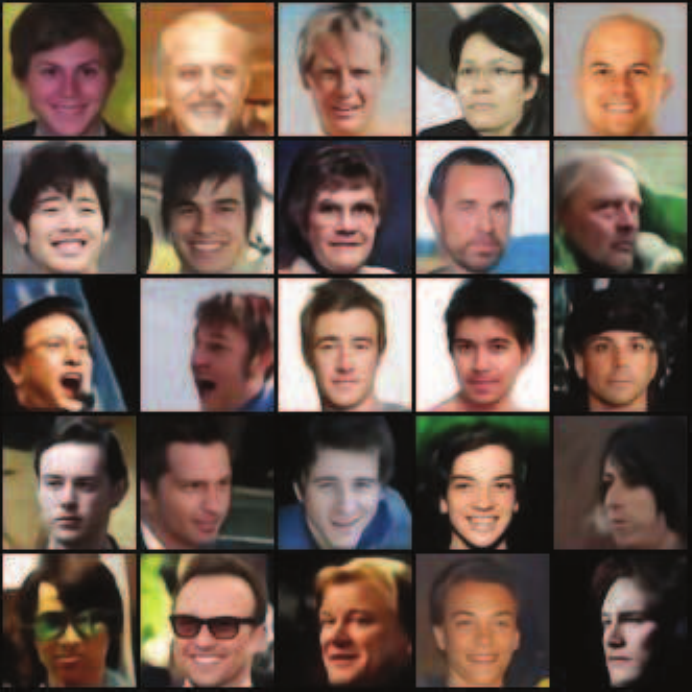}
		\vskip-0.1cm\centering {\footnotesize (c) Reconstructed image. Training SNR=5dB}
	\end{minipage}
	\caption{ The original image of TX $1$ and the reconstructed image under different training SNRs.}
	\label{fig10}
\end{figure*}

\begin{figure*}[!ht]
	\begin{minipage}[t]{0.32\textwidth}
		\centering
		\includegraphics[width=5cm]{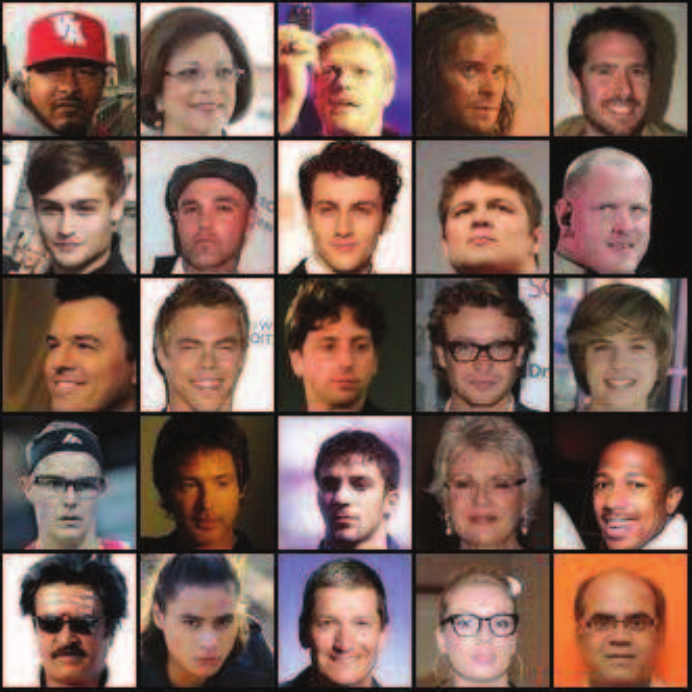}
		\vskip-0.1cm\centering {\footnotesize (a) Original image}
	\end{minipage}
	\begin{minipage}[t]{0.32\textwidth}
		\centering
		\includegraphics[width=5cm]{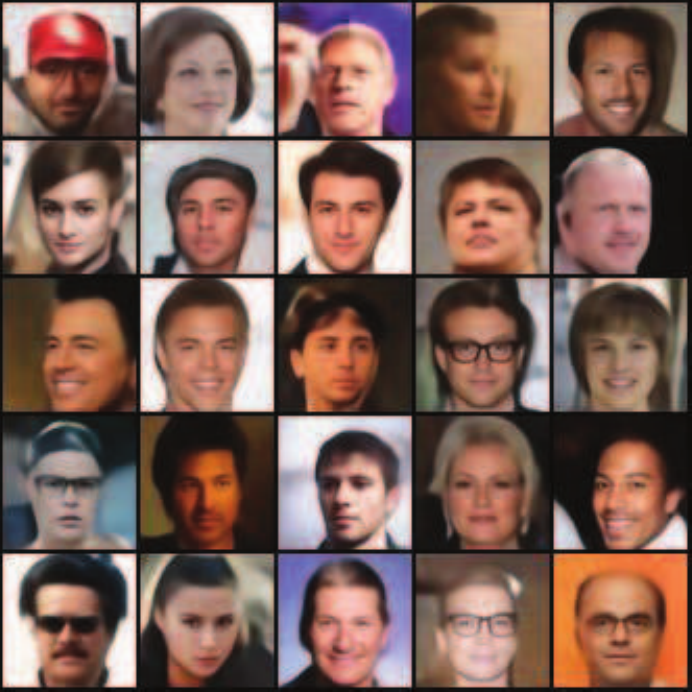}
		\vskip-0.1cm\centering {\footnotesize (b)  Reconstructed image. Training SNR=0dB}
	\end{minipage}
	\begin{minipage}[t]{0.32\textwidth}
		\centering
		\includegraphics[width=5cm]{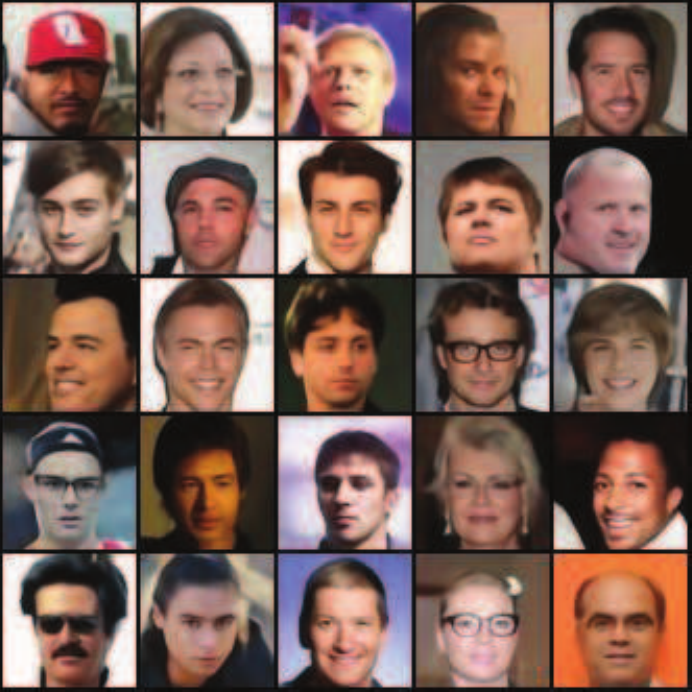}
		\vskip-0.1cm\centering {\footnotesize (c) Reconstructed image. Training SNR=5dB}
	\end{minipage}
	\caption{  The original image of TX $2$  and the reconstructed image under different training SNRs.}
	\label{fig11}
\end{figure*}

Fig. \ref{fig8} shows the performance of RXs decoding their own semantic features and the semantic features of the interfering TXs in the interference channel, where the input images of the two TXs are the same. As shown in Table \ref{table5},  the PSNR of  RX $1$ for decoding the  received signal ${\mathbf{y}_{1}}$ is $25.647$dB, and  the PSNR of  RX$1$ for decoding the  interference signal ${\mathbf{g}_{1,2}\mathbf{x}_2}$ is $7.350$dB.
 Moreover, the PSNR of  RX $2$ for decoding the   received signal ${\mathbf{y}_{2}}$ is $25.675$dB, and  the PSNR of  RX $2$ for decoding the  interference signal ${\mathbf{g}_{2,1}\mathbf{x}_1}$ is $7.469$dB.
 Furthermore, where Fig. \ref{fig8} (a) shows the input image, and Fig. \ref{fig8} (b), (c), (d) and (e) respectively show  the image of    RX $1$ decoding the received signal $\mathbf{y}_{1}$,    RX $1$ decoding the received interference  $\mathbf{x}_2$,  RX $2$ decoding the received signal $\mathbf{y}_{2}$, and    RX $2$ decoding the received interference ${\mathbf{x}_1}$, respectively.

Note that, for  the   same inputs, the RXs can only decode their own semantic features, and cannot decode the other semantic features, which  verifies the  approximate orthogonality of semantic features of the proposed SFDMA scheme.


Fig. \ref{fig9} compares  the image restoration  performance of the four   schemes   for image reconstruction   tasks on the CelebA  dataset  in Rayleigh channel with  (a) Training SNR=5dB and PSNR as  performance metric,  (b) Training SNR=5dB and MS-SSIM as  performance metric.
 Fig. \ref{fig9} shows that the SFDMA scheme over Rayleigh channel has higher image reconstruction capability than the existing deep JSCC  with interference at all SNRs, especially at low SNRs, the advantage of the SFDMA scheme is more prominent. Moreover, the SFDMA scheme is even better than  the deep JSCC without interference at low SNR, because  the SFDMA scheme is trained with interference and the Deep JSCC is trained without interference.

Fig. \ref{fig10} and Fig. \ref{fig11} show the reconstructed images of TX $1$ and TX $2$ at 0dB and 5dB,  respectively. It can be seen that under the condition of low SNR, the  proposed SFDMA method  can better reconstruct the original image.

\subsection{ Experimental results   of adaptive power control}

\begin{figure}[!ht]
	\begin{minipage}[t]{0.5\textwidth}
		\centering
		\includegraphics[width=7.5cm]{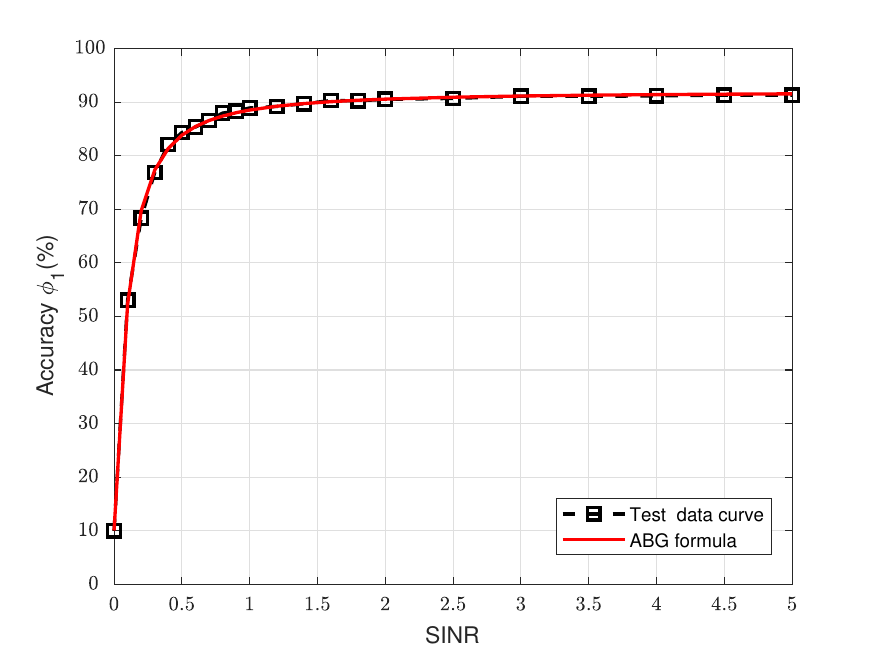}
		\vskip-0.1cm\centering {\footnotesize (a) Test data and  ABG formula  }
	\end{minipage}
	\begin{minipage}[t]{0.5\textwidth}
		\centering
		\includegraphics[width=7.5cm]{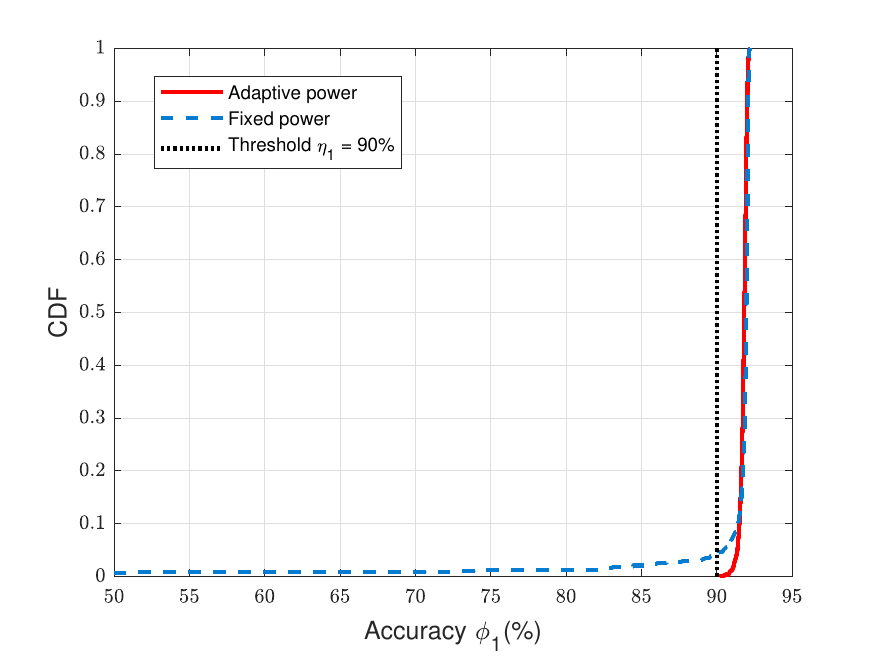}
		\vskip-0.1cm\centering {\footnotesize (b) ${\eta _1}=90\%$ }
	\end{minipage}
	\caption{(a)   Test data and  ABG formula,   (b) CDF of classification accuracy ${\phi _1}$ with threshold ${\eta _1}=90\%$ }
	\label{fig12}
\end{figure}

To verify the accuracy of   ABG formula,   the classification accuracy of the 1st transmission pair  versus  SINR is demonstrated in Fig. \ref{fig12} (a), where
the dashed lines with squares is plotted based on the test data of the SFDMA  networks, and the   solid line is  the     ABG formula curve.

 To quantitatively characterize the difference   between   ABG formula  and the test  data, we adopt  the  Adjusted R-Square to measure the  goodness of the fit.  Let $\varsigma_1$ denote the  adjusted R-Square of the  ABG formula.
 The closer the value of  $\varsigma_1$  is to 1, the better the fitting performance of ABG formula is.  Moreover,  the values of parameters in ABG formula $ \alpha_1$,  $\beta_1$,  $\gamma_1$, $\tau_1$ and   Adjusted R-Square $\varsigma_1$ are listed in Table \ref{table7}.

  	\begin{table}[htbp]
  	\caption{Parameters of ABG formula}
  	\label{table7}
  	\centering
	\scalebox{0.8}{
  	
  	\begin{tabular}{|c|c|c|c|c|c|}
  		\hline
  		\rule{0pt}{8pt}Parameters &  $ \alpha_1 $ &  $\beta_1$ &  $\gamma_1$ & $\tau_1$ &$\varsigma_1$  \\ \hline
  		
  		\rule{0pt}{6.5pt}  Value &  91.95 & 10.50 &  81.90 & 1.329 & 0.999\\ \hline		
  	\end{tabular}
}
  \end{table}

  As shown in Fig. \ref{fig12} (a),    ABG formula   can well fit the  classification accuracy performance of  SINR, and the goodness of fit $\varsigma_1$ is 0.999,  which  verify the accuracy of  the proposed  ABG formula.
    Moreover,    classification accuracy improves rapidly as SNR
increases, and then slowly increases until it reaches  the
upper bound.

 Fig. \ref{fig12} (b)  shows the cumulative distribution functions (CDFs)  of classification accuracy of the
  fixed transmitted power  method  and  that of the proposed adaptive power control method, where  the classification accuracy threshold  is $90\%$.
In Fig.\ref{fig12} (b), the fixed transmitted power   is 11.50dBm and the average power of the adaptive power control method is 10.43dBm. Fig. \ref{fig12} (b) shows that  the outage of the fixed transmitted power method is $5\%$, while  the outage  of proposed adaptive power control method is 0. This demonstrates that the adaptive power control method     can   effectively guarantee the QoS of semantic communications for random fading channels.

\section{Conclusions}

To address the massive access requirement under limited physical resources, we investigated a new semantic feature domain  by
utilizing DL, and proposed a   SFDMA scheme, where multiple users can transmit simultaneously in the same time-frequency resources. In our SFDMA scheme, the semantic encoder projects the semantic information of multiple users into distinguishable feature subspaces, where the discrete semantic feature representation   vectors of the users are approximately orthogonal to each other. Furthermore, for the multi-user semantic interference network with inference tasks, we propose a robust RIB semantic encoder and decoder based SFDMA, which achieves   approximate orthogonal transmission. Moreover, for the multi-user semantic interference network with image reconstruction tasks, we propose a multi-user JSCC scheme based on SFMDA, which protects the semantic information from being decoded by other users while realizing approximately orthogonal transmission of semantic features.
Furthermore,   the relationship between inference accuracy and transmission power are established, and the  adaptive power control methods with closed-form expression are derived for   inference tasks. Simulation results show
	that our proposed SFDMA scheme  can achieve  approximately orthogonal transmission of   semantic features, and outperforms existing approaches  for both classification tasks and image     reconstruction tasks.
	 Our proposed adaptive power control method   can         effectively guarantee the QoS  semantic communications in random fading channels.
 This paper designed a new MA for  multi-user digital
semantic  communication networks and proposed the first   theoretical   expression between inference accuracy and transmission power for  semantic communications.

\bibliographystyle{IEEEtran}
\bibliography{reference}

\end{document}